\documentclass[pre,twocolumn,superscriptaddress,floatfix]{revtex4-2}
\usepackage{amsmath,amssymb,dsfont,mathtools}
\usepackage[utf8]{inputenc}
\usepackage[T1]{fontenc}
\usepackage[english]{babel}

\newcommand\lbr{\left\lbrace}
\newcommand\rbr{\right\rbrace}
\newcommand\lp{\left(}
\newcommand\rp{\right)}
\newcommand\ls{\left[}
\newcommand\rs{\right]}

\makeatletter
\newcommand{\vast}{\bBigg@{4}}
\makeatother

\begin{document}
\title{Hypermodularity and community detection in hypergraphs}
\author{Charo I. \surname{del Genio}}
    \affiliation{Institute of Smart Agriculture for Safe and Functional Foods and Supplements, Trakia University, Stara Zagora 6000, Bulgaria}
    \affiliation{Research Institute of Interdisciplinary Intelligent Science, Ningbo University of Technology, 315104 Ningbo, China}
    \affiliation{School of Mathematics, North University of China, 030051 Taiyuan, China}

\date{\today}

\begin{abstract}
Numerous networked systems feature a structure of nontrivial
communities, which often correspond to their functional modules.
Such communities have been detected in real-world biological,
social and technological systems, as well as in synthetic models
thereof. While much effort has been devoted to developing methods
for community detection in traditional networks, the study of
community structure in networks with higher-order interactions
is still not as extensively explored. In this article, we introduce
a formalism for the hypermodularity of higher-order networks
that allows us to use spectral methods to detect community structures
in hypergraphs. We apply this approach to synthetic random networks
as well as to real-world data, showing that it produces results
that reflect the nature and the dynamics of the interactions
modelled, thereby constituting a valuable tool for the extraction
of hidden information from complex higher-order data sets.
\end{abstract}

\maketitle

\section{Introduction}
Many complex systems have the structure of a network,
in which pairs of elements of a discrete set, called
the nodes, are connected when they share a common property~\cite{Alb02,New03,Boc06,Boc14}.
Typically, such networks have structural features with
different characteristic length scales. The smallest
length scale is that of the number of links (also called
edges) of a given node, which takes the name of degree.
The links between the neighbours of a specific node define
the local neighbourhood structure, which is the next
larger length scale. As one considers the whole network,
one may find sets of nodes whose elements have more connections
amongst themselves than they have with nodes in other
such sets. When this happens, these mesoscale sets take
the name of communities, and the network is said to exhibit
a nontrivial community structure. The presence of network
communities has been detected in a large number of real-world
systems, such as protein networks, social networks,
food webs, and synthetic models of complex
systems of diverse nature~\cite{Pim79,Gar96,Fla02,Eri03,Kra03,Lus04,Gui05,Pal05,Hus07,For10}.
In general, the community structure of a network affects
the behaviour of the system, and often one can trace a
direct correspondence between communities and structural
units responsible for specific functions.

As a result, a large number of methods have been developed to detect
communities in networks~\cite{For10,Che19}. A particularly successful
approach is the maximization of a quality function that measures how
pronounced a partition of a network into communities is by computing
the difference between the density of the observed intracommunity links
and that of a suitably chosen null model. The most commonly used such
function is called \emph{modularity}~\cite{New06}. However, finding
the network partition that maximizes it is an NP-hard computational
problem~\cite{Bra08}. Thus, large efforts have been devoted to creating
algorithms that provide the best possible approximation to the global
maximum in a workable running time~\cite{Che04,Duc05,Blo08,Noa09,Sun09,Goo10,LeM11,Sob14,Tre15,Lu20}.

Recently, however, it has become increasingly clear that dyadic interactions
are only a part of the important relationships that occur in complex systems.
In fact, researchers have recognized the relevance of collective interactions
for the functioning of systems as diverse as social groups, neural networks
and multispecies assemblages of microorganisms, sparking a new interest in
structures with higher-order interactions~\cite{Bat20,Boc23}. As a result, the
main type of structural backbone that is currently generating considerable attention
is no longer that of a graph, but rather that of a hypergraph. In hypergraphs,
edges do not necessarily connect pairs of elements, but rather they can connect
triplets, quadruplets, or any other number of nodes~\cite{Ber73}.

Unfortunately, the landscape of research on the subject of community detection
in hypergraphs is not as well developed as in the case of traditional graphs.
However, several methods have been proposed in recent
years that address the problem from different directions. In some instances, these
approaches are dynamical in nature, and analyze the outcome of a stochastic process
on the hypergraph to infer its structure of modules and communitites~\cite{Eri22}.
Other methods, instead, rely on projecting the hypergraph onto a graph~\cite{Kum20},
or on creating an auxiliary graph on which to carry out a community-detection
procedure~\cite{Con23}. At the cost of a significant potential to lose information,
especially when performing projections~\cite{Boc23}, these techniques have the
advantage of reducing the dimensionality of the problem, and resulting, therefore,
in faster algorithms. A similar idea of dimensional reduction is behind methods
based on embedding, in which the starting hypergraph is related to a lower-dimensional
Euclidean space, and communities are detected by solving an optimization problem
or via the application of existing clustering algorithms~\cite{Wri19,Zhe23}. Such
embeddings can themselves be obtained via the application of spectral methods,
which have already proved extremely powerful for the task of finding communities
in classic networks~\cite{New06}. Indeed, spectral approaches often form the basis
of the more advanced higher-order methods, where they are occasionally integrated
by Bayesian inference algorithms~\cite{Ang15}. The already mentioned possibility
of information loss has also led to the development of algorithms that operate
directly on the whole hypergraph. In this case, tensor methods are necessary, as
one needs to relate the existence of communities to some property of the adjacency
tensor of the higher-order network. One possible approach is to use the idea of
tensor decomposition, which has been, in fact, long suggested as being applicable
to community detection~\cite{Rab17}, and has been eventually used for this goal,
once more in conjunction with a Bayesian framework~\cite{Ke19}. Alternatively,
others have exploited the spectral features of higher-order equivalents of the
graph Laplacian, effectively extending to higher-order structures the existing
knowledge of the relation between the Laplacians of traditional graphs and the
presence of communities~\cite{Kri21}. An exception to these approaches is an extension
to hypergraphs of the Chung-Lu model, which results in a modularity measure that
is closely related to the one originally used in traditional networks~\cite{Kam19,Kam24}.

Here, we introduce and characterize modularity for hypergraphs
using a formalism that allows us to create a spectral algorithm
for its maximization. Our method differs from the existing ones
mentioned above as it does not rely on embeddings of any kind
and does not require the structure of the higher-order network
to be a simplicial complex. Also, while the core of our approach
is a spectral method, this is applied onto a modularity tensor,
rather than on the adjacency tensor or the Hodge Laplacian of
the structure. Note that the specific form of the modularity measure
we use is mathematically equivalent to that of Refs.~\cite{Kam19}
and~\cite{Kam24}. However, we are able to operate a tensor decomposition
on it because of the different formalism adopted, which rests
on a derivation of a closed-form combinatorial expression that
enables us to cast the hypermodularity equation in vector form.
We then validate our method by analyzing random hypergraphs and
real-world higher-order networks, showing that it detects communities
that correspond to the similarity of cognitive processes in people
and reflect the underlying social dynamics.

\section{Results}
\subsection{Hypermodularity}
In a fully general case, a hypergraph may contain edges involving different
numbers of nodes. If all the edges have the same size~$k$, the hypergraph is
called $k$-uniform. Note that this means that classic networks with only dyadic
interactions are 2-uniform hypergraphs. However, each edge size is independent
from all the others, and they all correspond to different specific orders of
many-body interactions. Thus, a hypergraph~$H$ can be considered as the union
of multiple independent uniform sub-hypergraphs, each consisting of edges of
a single size. In turn, this consideration also applies to the task of finding
the community structure of~$H$, since, for example, a group of nodes may form
strong 3-body interactions but no significant pairwise ones. Therefore, in
the following, we study the problem on uniform hypergraphs, since a solution
on them also provides one to the general case.

The idea behind classic network modularity is to exploit
a measure based on the difference between the number of
observed edges between two nodes and the one that would
be expected if the placement of all the edges in the graph
were randomized while keeping the degrees of the nodes unchanged.
Given a partition of a network with $N$~nodes and $m$~edges
into a set of communities, this approach results in the
following expression for the modularity~$Q$~\cite{New06}:
\begin{equation}\label{modul}
 Q = \frac{1}{2m}\sum_{i=1}^N\sum_{j=1}^N\ls\lp A_{i,j} - \frac{d_i d_j}{2m}\rp\delta_{C_i, C_j}\rs\:,
\end{equation}
where~$\mathbf A$ is the adjacency matrix, $d_i$ is the degree
of node~$i$, $\delta$ is Kronecker's symbol and~$C_i$ is the community
to which node~$i$ has been assigned. Applying the same idea to
a $k$-uniform hypergraph, we introduce an expression for the \emph{hypermodularity}
of a partition:
\begin{multline}\label{hypermod}
 Q = \frac{1}{k!m}{\sum}_{\lbr v_1, \dotsc, v_k\rbr}\vast[\Bigg( A_{v_1, \dotsc, v_k} -\\
 \left.\left. \frac{\lp k-1\rp!}{\lp km\rp^{k-1}}\prod_{i=1}^k d_{v_i}\rp\prod_{\substack{i,j=1\\i\neq j}}^k \delta_{C_{v_i}, C_{v_j}}\rs\:.
\end{multline}
In the formula above, $\mathbf A$ is the adjacency
tensor of the hypergraph, the combinatorial factor
before the product of degrees accounts for the number
of available ways to join nodes of given degrees in
an edge, and the sum is on all the possible choices
of~$k$ nodes, allowing for repetitions. Thus, the
expression directly measures the difference between
the number of hyperedges of size~$k$ that are found
amongst the nodes in each community and the number
that would be expected if the same hyperedges had
been placed at random. The task is thus to find the
partition of the network that maximizes~$Q$. Note
that $\mathbf A$ is more precisely described as a
hypermatrix, rather than a tensor, since it does not
represent a multilinear application. However, with
a slight abuse of notation we will still refer to
it as a tensor, as this is the most commonly adopted
terminology, especially in the applied literature.
Also note that, for $k=2$, the expression reduces
to that of graph modularity, which becomes therefore
a special case of this more general formulation, as
desired.

The principal difference between the hypermodularity
of Eq.~\eqref{hypermod} and the classic modularity
for graphs is that the latter can be cast
into a pure matrix equation, whereas such an expression
cannot be obtained straightforwardly in the hypergraph
case. Thus, one cannot directly use the same methods
employed on traditional networks to produce a bipartition,
which is a starting step in many community detection
algorithms.

\subsection{Spectral partitioning}\label{specpar}
To solve this problem, first note that if the hypergraph
is simple, meaning that it does not have multiple edges,
as it is often the case in real-world structures, then one
can write $A_{v_1, \dotsc, v_k} = \mathbf{1}_{\lp v_1, \dotsc, v_k\rp\in H}$,
where~$\mathbf 1$ is an indicator variable whose value is~1
if the edge $\lp v_1, \dotsc, v_k\rp$ belongs to~$H$
and~0 otherwise. Thus, Eq.~\eqref{hypermod} can be rewritten
as
\begin{multline}\label{hypermod2}
 Q = \frac{1}{k!m}{\sum}_{\lbr v_1, \dotsc, v_k\rbr}\vast[\Bigg( \mathbf{1}_{\lp v_1, \dotsc, v_k\rp\in H}\\
 - \left.\left.\frac{\lp k-1\rp!}{\lp km\rp^{k-1}}\prod_{i=1}^k d_{v_i}\rp\prod_{\substack{i,j=1\\i\neq j}}^k \delta_{C_{v_i}, C_{v_j}}\rs\:.
\end{multline}
But then, one can consider the whole term within the round brackets
in the equation above as a hypersymmetric data tensor~$\mathbf B$, which,
in principle, can be analyzed using higher-order singular-value decomposition~(SVD),
whose applications to data science and machine learning have proved
to be very fruitful~\cite{Raj13,San15,Tag17,Tag17_2,Vil23}. Given a
data tensor, its higher-order SVD is a compact SVD carried out on one
of its standard flattenings. These are dimensional reductions of the
tensor that compact all its dimensions, except for an arbitrarily chosen
one, into a single dimension. In our case, this means that the standard
flattenings of~$\mathbf B$ are the matrices~$\mathbf E^{(i)}$ defined
by
\begin{multline}\label{manyflattenings}
 E^{(i)}_{v_i, \lp v_1, v_2, \dotsc, v_{i-1}, v_{i+1}, \dotsc, v_k \rp} = \mathbf{1}_{\lp v_1, \dotsc, v_k\rp\in H}\\ - \frac{\lp k-1\rp!}{\lp km\rp^{k-1}}\prod_{j=1}^k d_{v_j}\:,
\end{multline}
where~$i$ is any integer from~1 to~$k$. Note that, effectively,
the column index of~$\mathbf E^{(i)}$ is a sort of collective index,
which can take a number of values that is equal to the number of
all possible combinations of allowed values for all the indices
of~$\mathbf B$ except the $i$-th one. Also note that, with these
manipulations, we do not neglect or discard any information that
is present in the initial equation.

In general, the compact SVD of an $N\times M$ matrix~$\mathbf X$
provides three matrices~$\mathbf U$, $\boldsymbol\Sigma$ and~$\mathbf V$
such that $\mathbf X=\mathbf U\boldsymbol\Sigma\mathbf V^\dagger$,
where~$\dagger$ indicates Hermitian conjugation. More specifically,
$\mathbf U$ and $\boldsymbol\Sigma$ are $N\times N$ and~$\mathbf V$
is $M\times N$. Also, the columns of~$\mathbf U$ and those of~$\mathbf V$
are the left singular vectors and the right singular vectors of~$\mathbf X$,
respectively, whereas $\boldsymbol\Sigma$ is diagonal and it contains
the (positive) singular values of~$\mathbf X$. For our purposes,
the procedure is useful because, since~$\mathbf B$ is real, so are
its flattenings~$\mathbf E^{(i)}$. This means that the Hermitian
conjugation reduces to transposition, and that the matrix~$\mathbf U$
is guaranteed to be real and orthogonal. Thus, if we could establish
a relation between the product of Kronecker deltas in Eq.~\eqref{hypermod2}
and some vector quantities encoding the assignments of the nodes
into communities, we would be able to maximize the hypermodularity
by imposing the partition described by the left singular vector
corresponding to the largest singular value.

To find this vector, one can use a variety of methods.
However, since we are only interested in the first left
singular vector, we can use an equivalent of the power
iteration, thus avoiding the need to perform a full
decomposition.
The procedure is as follows. First, create an initial
$N$-dimensional vector~$\mathbf{z_0}$ whose components
are standard normal random variables. Normalize~$\mathbf{z_0}$,
and compute the vector $\mathbf{z_1} = \mathbf E^{(i)}\lp{\mathbf E^{(i)}}^{\mathrm T}\mathbf{z_0}\rp$,
where~T indicates transposition. Normalize~$\mathbf{z_1}$
and repeat the multiplication to obtain~$\mathbf{z_2}$.
After a large number of iterations, the resulting vector~$\mathbf{z_\infty}$
is the first left singular vector of~$\mathbf E^{(i)}$.
Then, one can determine the community assignment based
on the signs of the elements of the vector: if the $i$th
element is positive, assign node~$i$ to the first community;
if it is negative, assign it to the second one.

Note that, in principle, Eq.~\eqref{manyflattenings}
defines~$k$ different possible flattenings of the data
tensor. Thus, one may wonder which one of them to use
for the~SVD. However, since the tensor is hypersymmetric,
one can intuitively see that any dimension can be chosen
for the flattening, because reordering them does not change
the tensor. Thus, in the following we use the $k$-th one
without loss of generality, omitting from the formulae
the explicit mention of the dimension used. For a more
formal proof of this, see Appendix~\ref{flatident}.

The next task on the road to building a method
for spectral bisection of a higher-order network
is finding a way to express Eq.~\eqref{hypermod2}
in terms of vector quantities. To start, recall
that one can write the simple Kronecker delta
of Eq.~\eqref{modul} as
\begin{equation}\label{deltas}
 \delta_{C_i, C_j} = \frac{1}{2}\lp s_i s_j + 1\rp\:,
\end{equation}
where~$s_i$ are spin-like community-identifying
variables that can take the values~$+1$ and~$-1$,
depending on the community to which node~$i$ is
assigned. Next, note that, given $k$~nodes, the
product of deltas in Eq.~\eqref{hypermod2} does
not need to include all $\binom{k}{2}$ possible
combinations thereof. In fact, because of the transitivity
of equality, the product is best expressed as a
chain, so that
\begin{equation}
 \prod_{\substack{i,j=1\\i\neq j}}^k \delta_{C_{v_i}, C_{v_j}} = \delta_{C_{v_1}, C_{v_2}} \delta_{C_{v_2}, C_{v_3}} \dotsb \delta_{C_{v_{k-1}}, C_{v_k}}\:.
\end{equation}

Thus, one can use Eq.~\eqref{deltas} to explicitly rewrite
the product chain for different values of~$k$. Then, exploiting
the facts that the square of any spin-like variable is equal
to~1 and that the sum of all the elements of the data tensor
vanishes by construction, one can find a general expression
for the hypermodularity of a bipartition, namely
\begin{equation}\label{vechypmod}
 Q = \frac{1}{2^{k-1}k!m}\mathbf s^{\mathrm T}\mathbf E\boldsymbol\sigma^{(k)}\:,
\end{equation}
where~$\mathbf s$ is a vector containing
the community assignments of all nodes and
$\boldsymbol\sigma^{(k)}$ is a vector with
$N^{k-1}$~components that can be written
as
\begin{multline}
 \boldsymbol\sigma^{(k)} = \lp\sigma_{1,1,\dotsc,1}, \sigma_{2,1,\dotsc,1}, \sigma_{3,1,\dotsc,1}, \dotsc, \sigma_{N,1,\dotsc,1},\right.\\
 \left.\sigma_{1,2,\dotsc,1}, \sigma_{2,2,\dotsc,1}, \sigma_{3,2,\dotsc,1}, \dotsc, \sigma_{N,N,\dotsc,N}\rp^\mathrm{T}\:.
\end{multline}
Here, the elements of~$\boldsymbol\sigma^{(k)}$
are characterized by $k-1$ indices whose sequence
is in inverse lexicographic order, and their actual
values are given by
\begin{widetext}
\begin{equation}\label{combvec}
  \sigma^{(k)}_{\xi_1,\xi_2,\dotsc,\xi_{k-1}} = \sum_{\substack{r=1\\r\text{ odd}}}^{k-1}\left.\sum_{\substack{\text{ordered $r$-choices }\lbr\varsigma_1,\varsigma_2,\dotsc,\varsigma_r\rbr\\\text{from }\lbr\zeta_1,\zeta_2,\dotsc,\zeta_{k-1}\rbr}}\alpha_1\varsigma_1\varsigma_2\dotsb\varsigma_r\right|_{\zeta_1=s_{\xi_1}, \zeta_2=s_{\xi_2}, \dotsc, \zeta_{k-1}=s_{\xi_{k-1}}}\:,
\end{equation}
\end{widetext}
where~$\alpha_1$ is the ordinal index
of~$\varsigma_1$, so that $\alpha_1=1$
if $\varsigma_1=\zeta_1$, $\alpha_1=2$
if $\varsigma_1=\zeta_2$, and so on.
For a full derivation of these expressions,
see Appendix~\ref{vectoreq}.

This treatment effectively transforms the original
tensor equation of Eq.~\eqref{hypermod} into a singular-value
problem on a matrix, without neglecting any hyperedge
present in the original network, and allowing one
to find the best bisection of a higher-order network.
However, it is clear that the best partition, in general,
may not consist of just two communities.

A natural idea is then to apply the same method to the communities
resulting from an initial partitioning, checking whether hypermodularity
increases when splitting those even further. However, care must be
taken when doing so. The reason is that a key step in the derivation
of Eqs.~\eqref{vechypmod} and~\eqref{combvec} relies on the vanishing
of the sum of all elements of~$\mathbf B$, as detailed in Appendix~\ref{vectoreq}.
However, when bisecting an already existing community, this condition
is no longer true, as one works with the subtensor corresponding to
the nodes in the community, rather than with the data tensor of the
entire network.

To find the correct way to deal with this situation,
consider the change in hypermodularity that would result
from a further split of a community~$C$. This can be
expressed as the sum of a positive term coming from
the new split and a negative one, which is the contribution
of the nodes in the community to the current value of
the hypermodularity. From Eq.~\eqref{hypermod2}, it is
clear that the former term consists of all and only
the elements of the subtensor of~$C$ that correspond
to nodes assigned to the same community. Similarly,
the latter term is the sum of all the elements of the
same subtensor. Thus, it is
\begin{multline}\label{deltaQ}
 \Delta Q=\frac{1}{k!m}\lp\sum_{\text{$k$-sets in $C$}}B_{v_1,\dotsc,v_k}\delta_{C_{v_1},C_{v_2}}\dotsb\delta_{C_{v_{k-1}},C_{v_k}}\right.\\
 \left.- \sum_{\text{$k$-sets in $C$}}B_{v_1,\dotsc,v_k}\rp\:.
\end{multline}
Using the same expression for the product of deltas
as before, but keeping the constant term that can no
longer be neglected, results in an expression that
can ultimately be manipulated in a way that still separates
the variables into a vector equation of the form
\begin{equation}\label{repbis}
 \Delta Q = \frac{1}{2^{k-1}k!m}\mathbf s^{\mathrm T}\mathbf{E'}\boldsymbol\sigma^{(k)}\:,
\end{equation}
where~$\mathbf{E'}$ is the flattening of a modified
community subtensor~$\mathbf{B'}$, obtained by adding
a correction to the elements on the main hyperdiagonal:
\begin{widetext}
 \begin{equation}\label{correq}
  B'_{v_1,\dotsc,v_k} = \begin{cases}
                        B_{v_1,\dotsc,v_k} - {\sum}_{s_{w_1}}\dotsb{\sum}_{s_{w_{k-1}}}B_{w_1,\dotsc,w_{k-1},v_k} & \quad\text{if $v_1=v_2=\dotsb=v_k$}\\
                        B_{v_1,\dotsc,v_k} & \quad\text{otherwise.}
                       \end{cases}
 \end{equation}
\end{widetext}
For a formal derivation of the equation
above, see Appendix~\ref{repbisec}. Note
that the correction terms vanish if one
considers the whole network, as expected.
Thus, one can always directly use Eq.~\eqref{repbis},
since it reduces to Eq.~\eqref{vechypmod}
when the network has not yet been partitioned.

Using the formalism just described,
we build an algorithm that searches
for the best partition of a $k$-uniform
higher-order network into communities
by carrying out repeated bisections.
Following the approach of Refs.~\cite{Tre15}
and~\cite{Bot16}, we also include refinement
steps, which we briefly describe below.

\subsection{Refinement steps}
After operating a bisection, we carry out
a first Kernighan-Lin-type refinement~\cite{Ker70}
by computing how the hypermodularity would
change if one were to switch the placement
of each node from the community to which
it was assigned to the other one.
All the
nodes are considered, and then, in a greedy
fashion, the best move is accepted. The remaining
nodes are then considered again, and the
best move is accepted each time, until all
the nodes have been evaluated. The best middle
point throughout the procedure is then found.
If it corresponds to an increase in hypermodularity,
the moves up to that point are permanently
accepted; otherwise, the state of the network
is reverted to the one at the start of the
procedure. The refinement is then repeated
until no further improvement is obtained.

\begin{figure*}[t]
\centering
\includegraphics[width=0.45\textwidth]{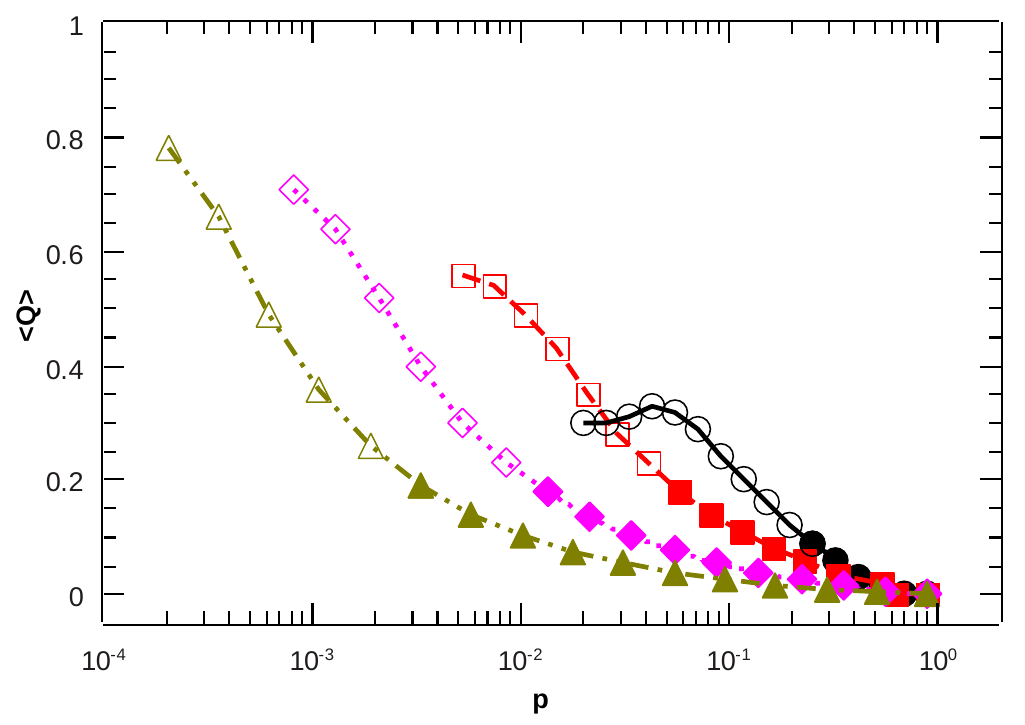}
\includegraphics[width=0.45\textwidth]{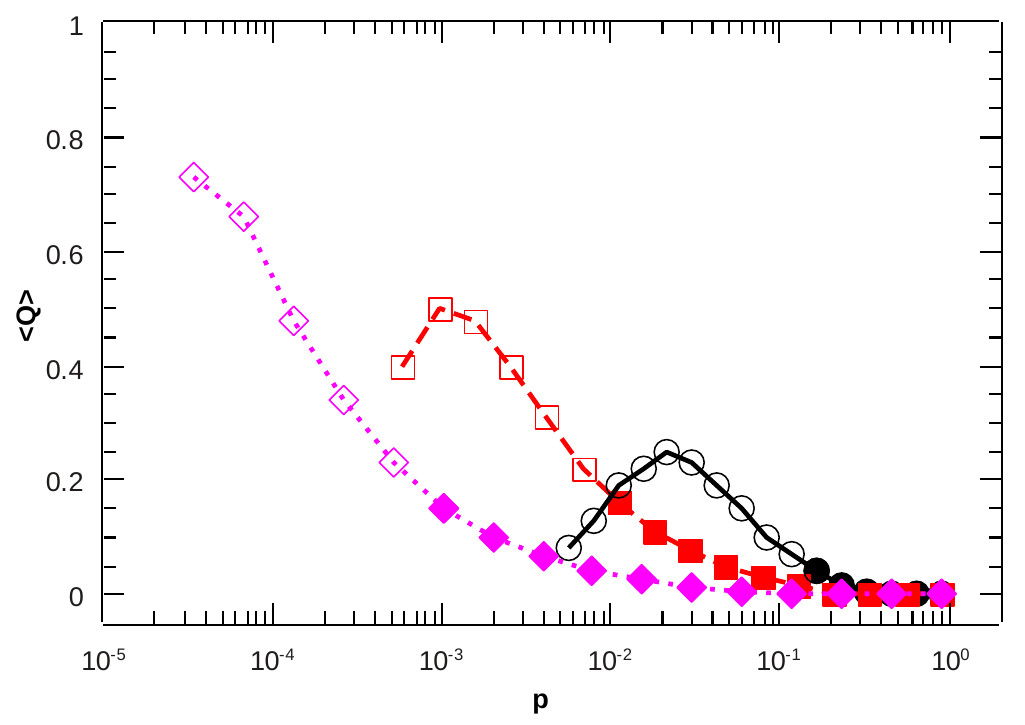}
\includegraphics[width=0.45\textwidth]{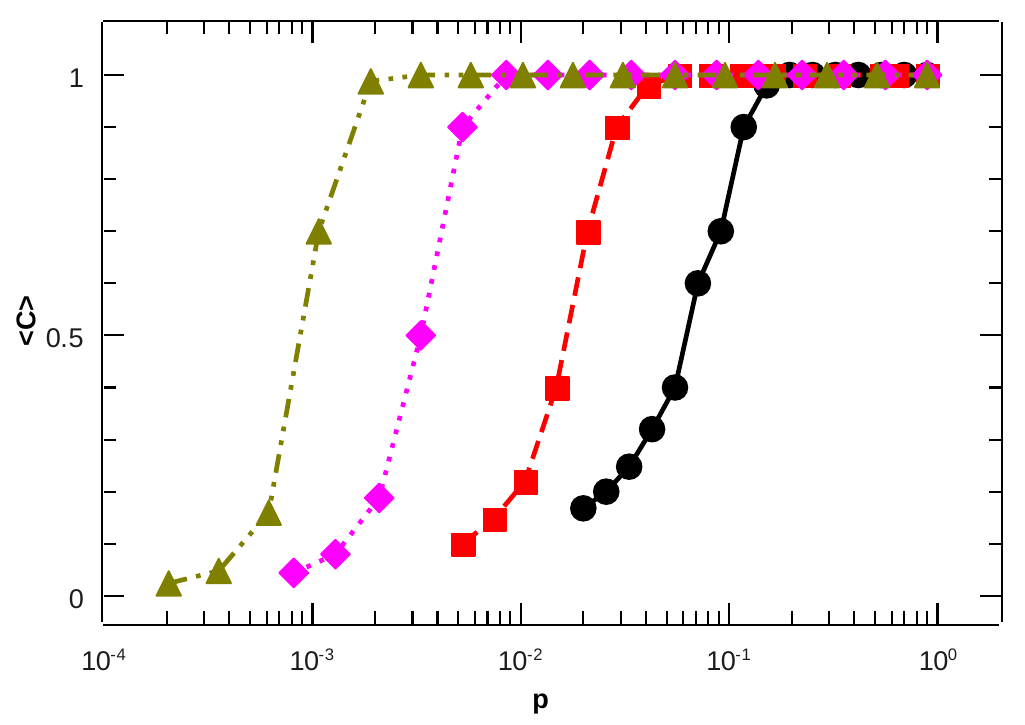}
\includegraphics[width=0.45\textwidth]{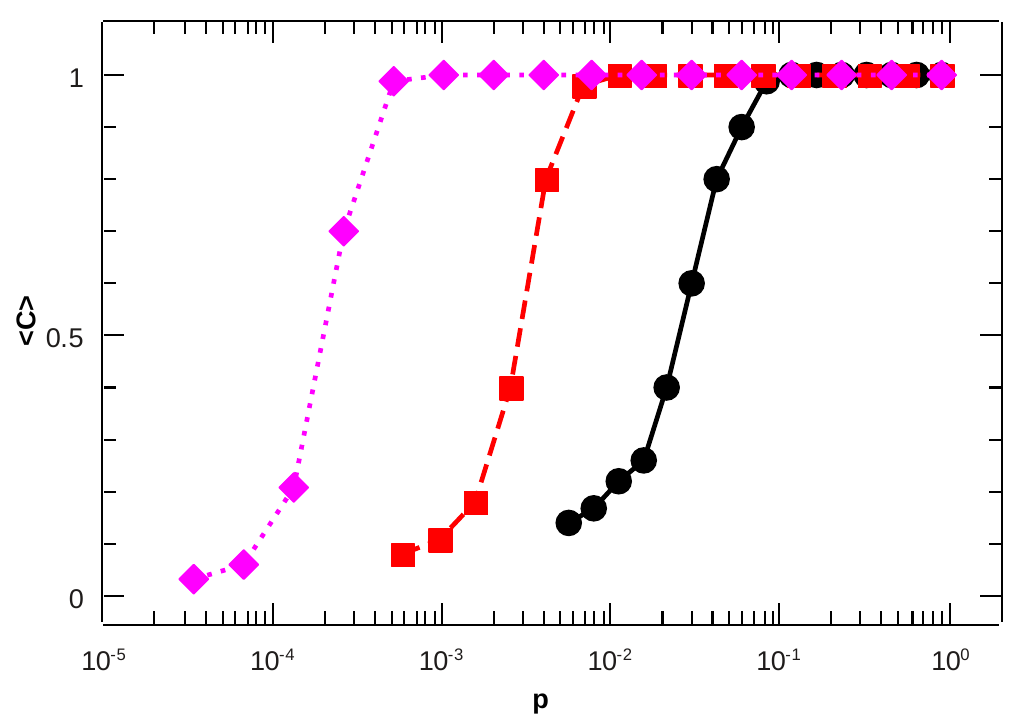}
\caption{\textbf{Maximum hypermodularity in random hypergraphs
is never above~$\mathbf{0.2}$.} So long as the networks remain
connected, the average maximum hypermodularity~$\left\langle Q\right\rangle$
of random $k$-uniform hypergraphs remains below~$0.2$, as shown
here for $k=3$ (left-hand top panel) and $k=4$ (right-hand top
panel). As the connection probability~$p$ decreases, the networks
start to fragment, and the value of the maximum hypermodularity
spuriously increases (empty symbols). The bottom panels show the
average fraction of nodes in the components of the networks~$\left\langle C\right\rangle$.
For all panels, black circles and solid lines correspond to $N=10$,
red squares and dashed lines to $N=20$, purple diamonds and dotted
lines to $N=50$ and green triangles and dotted-and-dashed lines
to $N=100$.}\label{hER}
\end{figure*}
To compute the change in hypermodularity
that each move would yield, first note that
changing the assignment of a single node~$i$
results in a single element of the vector~$\mathbf s$
switching sign. Similarly, the vector~$\boldsymbol\sigma^{(k)}$
only changes in the elements that correspond
to combinations of \mbox{$k-1$~nodes} involving
node~$i$. Thus, we can write $\boldsymbol\sigma^{(k)}\rightarrow\boldsymbol\sigma'^{(k)}=\boldsymbol\sigma^{(k)}+\boldsymbol\Delta\boldsymbol\sigma^{(k)}$
and $\mathbf s\rightarrow\mathbf s'=\mathbf s+\boldsymbol\Delta\mathbf s$,
where all the elements of~$\boldsymbol\Delta\mathbf s$
are~0, except for the $i$-th one, which
is equal to~$-2s_i$. Then, write the hypermodularity
of the partition after the change as
\begin{equation}
 \begin{split}
  Q' &= \frac{1}{2^{k-1}k!m}\lp\mathbf s+\boldsymbol\Delta\mathbf s\rp^\mathrm{T}\mathbf{E'}\lp\boldsymbol\sigma^{(k)} + \boldsymbol\Delta\boldsymbol\sigma^{(k)}\rp\\
  &= \frac{1}{2^{k-1}k!m}\lp \mathbf s^\mathrm{T}\mathbf{E'}\boldsymbol\sigma^{(k)} + \mathbf s^\mathrm{T}\mathbf{E'}\boldsymbol\Delta\boldsymbol\sigma^{(k)} \right. \\
  &\quad \left. + {\boldsymbol\Delta\mathbf s}^\mathrm T\mathbf{E'}\boldsymbol\sigma^{(k)} + {\boldsymbol\Delta\mathbf s}^\mathrm T\mathbf{E'}\boldsymbol\Delta\boldsymbol\sigma^{(k)}\rp\:.
 \end{split}
\end{equation}
From this equation, it follows that
\begin{equation}
 \begin{split}
  \Delta Q &= \frac{1}{2^{k-1}k!m}\lp \mathbf s^\mathrm{T}\mathbf{E'}\boldsymbol\Delta\boldsymbol\sigma^{(k)} + {\boldsymbol\Delta\mathbf s}^\mathrm T\mathbf{E'}\boldsymbol\sigma^{(k)}\right.\\
  &\quad \left. + {\boldsymbol\Delta\mathbf s}^\mathrm T\mathbf{E'}\boldsymbol\Delta\boldsymbol\sigma^{(k)}\rp\\
  &= \frac{1}{2^{k-1}k!m}\lp \mathbf{s'}^\mathrm{T}\mathbf{E'}\boldsymbol\Delta\boldsymbol\sigma^{(k)} + {\boldsymbol\Delta\mathbf s}^\mathrm T\mathbf{E'}\boldsymbol\sigma^{(k)}\rp\:.
 \end{split}
\end{equation}
This form is particularly convenient,
because the vector~$\boldsymbol\sigma^{(k)}$
is already known from the previous step.
Thus, the second term can be readily
computed as the product of the $i$-th
row of~$\mathbf{E'}$ and~$\boldsymbol\sigma^{(k)}$,
multiplied by $-2s_i$, and the first
term can be calculated similarly, by
accounting for all and only the combinations
of factors that include node~$i$.

After all existing communities have been evaluated
for further splits, and the local node-level refinement
has been considered, we perform a global node-level
optimization. The structure of this step is again
that of an iterated greedy algorithm that evaluates
the change in hypermodularity resulting from moving
each node from the community to which it is
currently assigned to any other possible community.
As before, the best move is accepted, and the evaluation
is repeated for the remaining nodes, until all nodes
have been moved. Then, the middle point in the sequence
of moves corresponding to the largest change in hypermodularity
is found, and all the moves up to that point are
permanently accepted if such change is positive. The
step is repeated until it produces an improvement in
hypermodularity.

The change in hypermodularity caused by moving
node~$i$ to community~$\alpha$ consists of two
contributions. The first is a negative one, which
is due to all and only the combinations of $k$~nodes
in the original community of~$i$ such that at
least one of them is node~$i$ itself. The second
is a positive change that is due to all and only
the combinations of $k$-nodes in community~$\alpha$,
with the temporary inclusion of node~$i$ within
it, such that at least one of them is node~$i$
itself. Both are sums of the elements of~$\mathbf B$
corresponding to the combinations of nodes taken,
divided by~$k!m$, and with the sign of the negative
contribution changed.

Finally, if the current partition consists
of more than one community, a community-level
optimization is carried out. As for the previous
two steps, also this is an iterated greedy
algorithm, repeated so long as the value of
hypermodularity increases. The moves considered
are the mergings of pairs of communities, akin
to the main procedure of the Louvain algorithm~\cite{Blo08}.
Merging community~$\alpha$ and community~$\beta$
causes a change in hypermodularity that is
due to all and only the combinations of $k$~nodes
such that at least one is in community~$\alpha$
and one is in community~$\beta$. As in the
previous step, each combination contributes
an amount equal to the corresponding element
of~$\mathbf B$ divided by~$k!m$. This means
that the potential changes in hypermodularity
can be represented as a matrix whose dimension
is the current number of communities. When
a move is accepted, it is convenient to take
all the nodes belonging to the community with
the higher index and assign them to the other
one. This way, for all subsequent moves, the
only matrix elements that change are those
involving the community with the lower index,
and, because of symmetry, only half a row
and half a column need to be updated.

An example of the functioning of our method
on a small 3-uniform hypergraph with 12~nodes
is provided in Appendix~\ref{example}.

\subsection{Numerical validation}
To demonstrate the applicability of the method,
we implemented the algorithm described above, and
analyzed both synthetic and real-world higher-order
networks. We started with the study of ensembles
of Erdős-Rényi-like random $k$-uniform hypergraphs,
to determine the behaviour of hypermodularity on
networks without a well-defined community structure.
In these networks, every possible hyperedge of
size~$k$ exists independently with the same probability~$p$.
We ran the algorithm on networks of size~10, 20,
50 and 100, with $k=3$ and $k=4$. The behaviour
of the average maximum hypermodularity as a function
of~$p$, shown in Fig.~\ref{hER}, shows that, on
connected networks, the best partition never corresponds,
on average, to a hypermodularity greater than~$0.2$.
However, as the connection probability decreases,
and the networks start fragmenting into separate
connected components, the maximum value of the
hypermodularity increases even as high as~$0.8$.
At first, this may seem evidence of a problem with
the measure, as there appear to exist partitions
of completely random networks whose hypermodularity
is close to the theoretical maximum of~1. However,
such values are actually spurious and they are,
in fact, due to the fragmentation of the networks.

\begin{figure*}[t]
\centering
\includegraphics[width=0.45\textwidth]{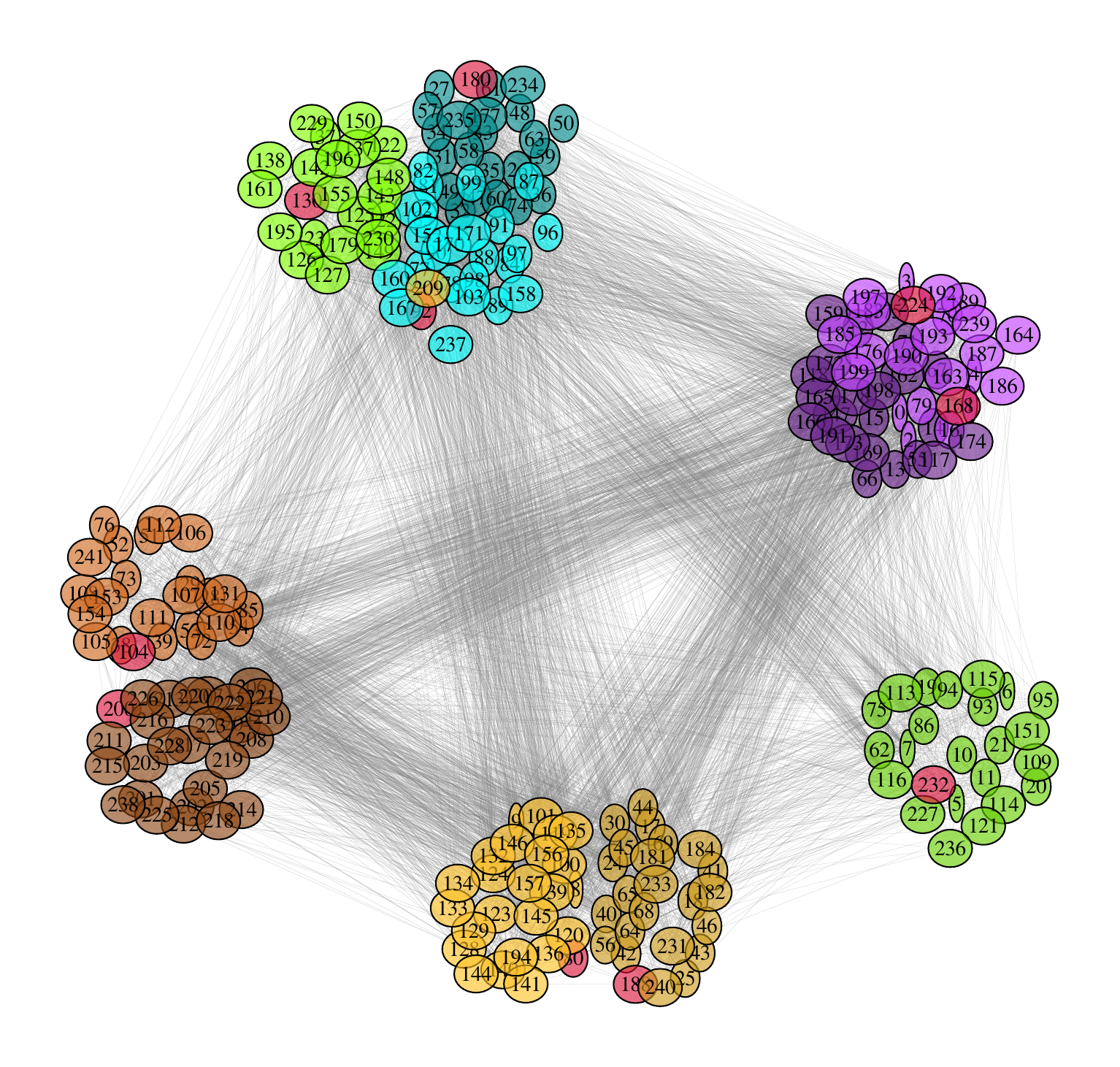}$\qquad\qquad\quad$
\includegraphics[width=0.45\textwidth]{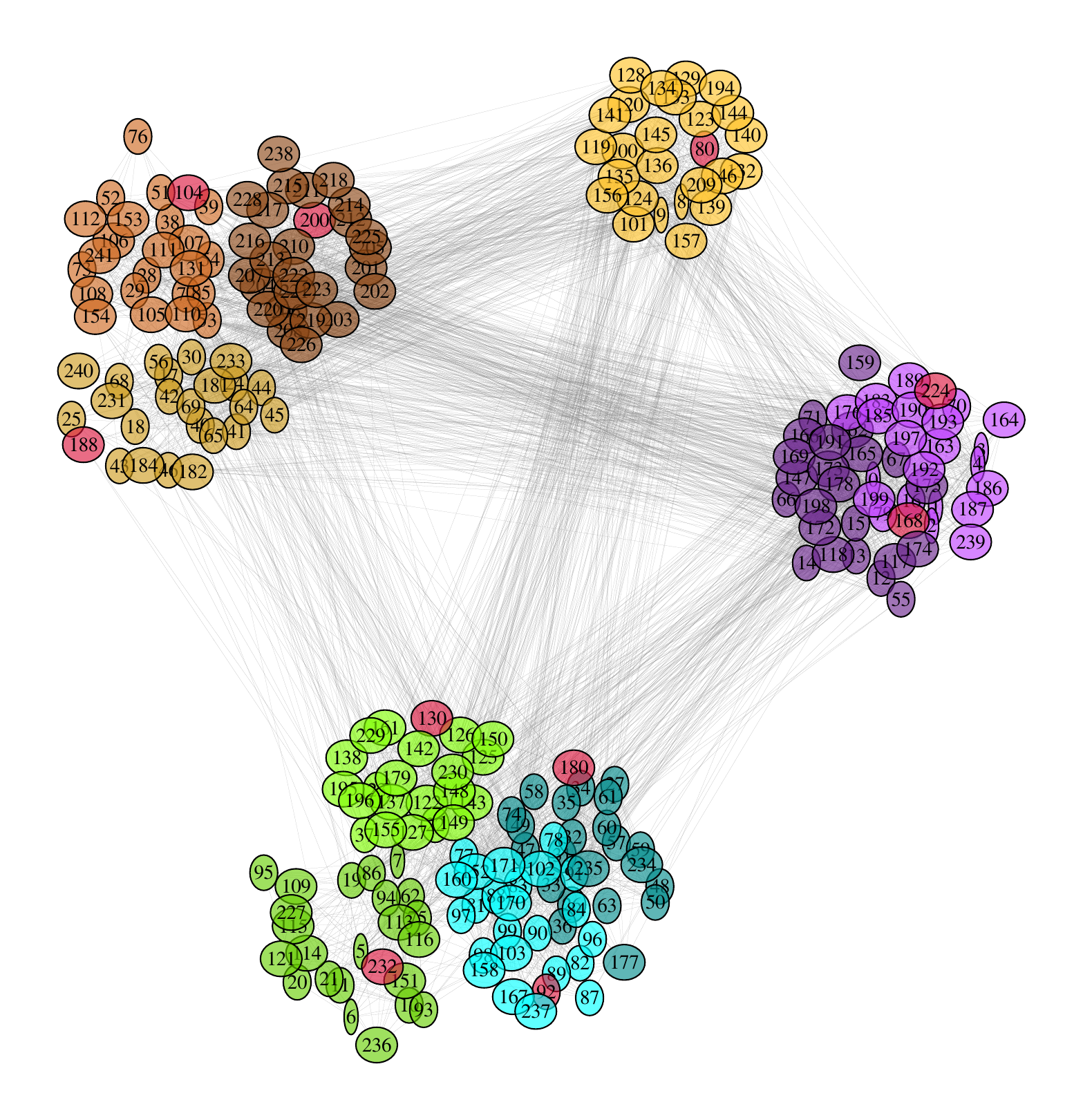}
\includegraphics[width=0.45\textwidth]{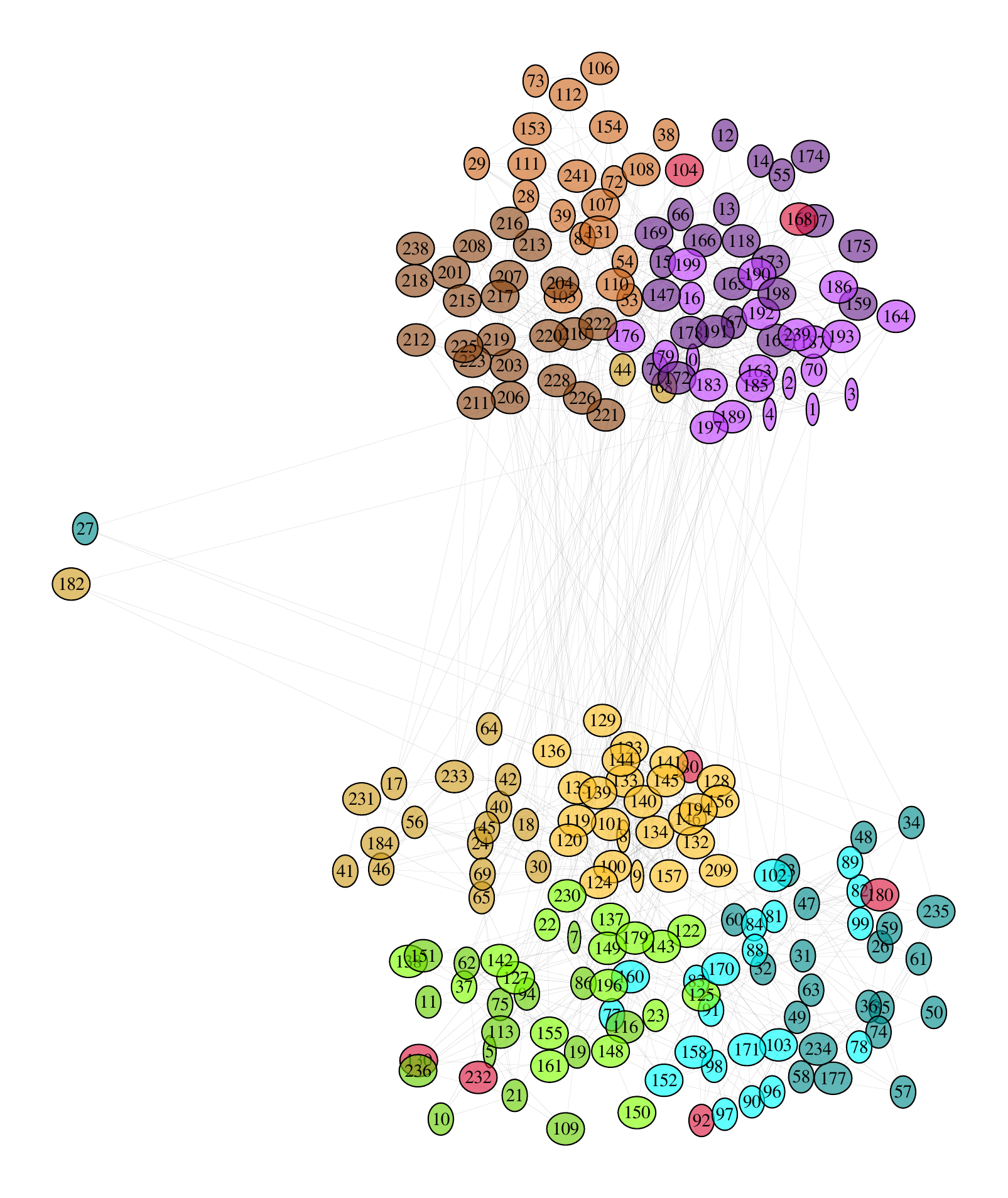}$\qquad\qquad\qquad\qquad\qquad\qquad\qquad\qquad$
\includegraphics[width=0.25\textwidth]{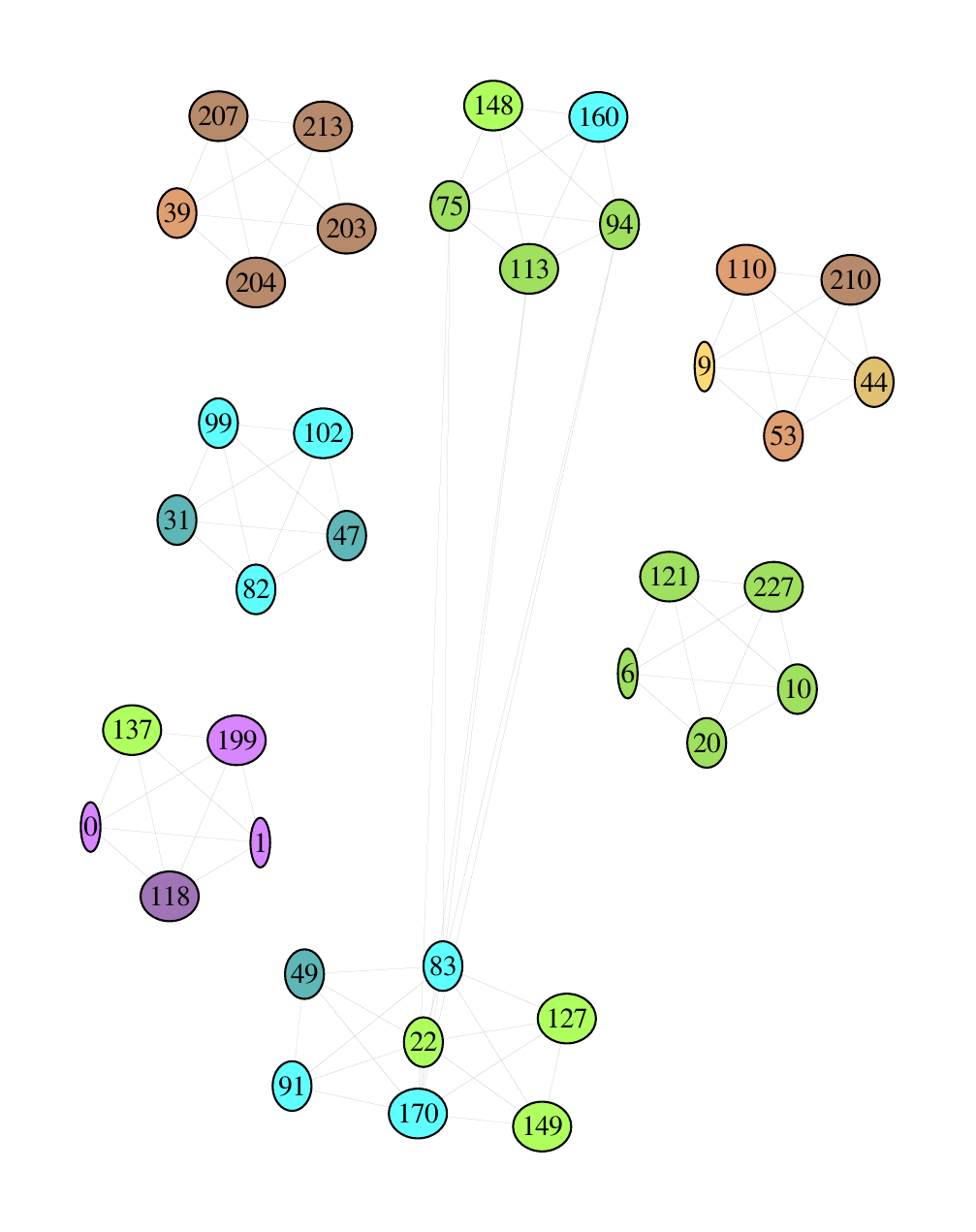}
\caption{\textbf{Higher-order communities of primary-school pupils
reveal underlying social dynamics.} The communities of 2-people and
3-people interactions (top-left and top-right panels, respectively)
show that tight-group interactions between primary-school pupils mostly
occur within age groups. Communities of looser groups of 4~people
(bottom-left) blur this line, with younger siblings being accepted
within groups of older children. Finally, interactions of 5~pupils
at a time (bottom-right) are rare, and possibly stronger, thereby
restoring the age separation seen for $k=2$ and $k=3$. In all panels,
yellow nodes correspond to pupils in year~1, brown nodes to pupils
in year~2, violet nodes to pupils in year~3, green nodes to pupils
in year~4, blue nodes to pupils in year~5 and red nodes to teachers.
Each year consists of two separate classes, signified by the shade
of the colour. Note that in preparing the figure, hyperedges ($k>2$)
have been projected onto cliques for ease of visualization.}\label{PriSch}
\end{figure*}
To understand how this happens, consider a limit
case of a hypergraph formed by~$m$ isolated hyperedges.
Its maximum hypermodularity, which is found by assigning
all the nodes of each edge to a separate community,
is $1-\frac{\lp k-1\rp !}{\lp km\rp^{k-1}}$, which,
for large~$m$, is approximately~1. However, the
hypermodularity of the individual components,
which consist of a single edge, is identically~0.
This shows that one must be very careful in using
hypermodularity on networks that are not connected,
and that in such cases the right approach is
to treat each component as an independent network.
Thus, our results also indicate that concerns of overfitting,
as have been occasionally voiced also for traditional
modularity, are mostly unfounded, and one can
take the value of~$0.2$ as a sort of physiological
expected maximum hypermodularity of a random
network, due to the non-vanishing probability
that, even in a random structure, some sets of
nodes may be more densely linked internally than
they are between each other.

We then tested the method on real-world data sets,
using contact information from a primary school~\cite{Ste11,Cho21}
and from a high school~\cite{Cho21,Mas15}. In the
former case, the pupils are classified according to
their grade, with two separate classes running for
each year from~1 to~5. The teachers, one per class,
were also involved in the study. The high-school
students, instead, are classified according to the
topic of the classes they were following. Specifically,
the data set contains three classes of biology, three
of mathematics and physics, two of chemistry and one
of psychology. In both cases, we detected the community
structure for edges of sizes~2 to~5, since no edges
of larger size were recorded.

The primary school results, shown in Fig.~\ref{PriSch},
are quite revealing. When we use $k=2$, which corresponds
to a traditional network, the algorithm divides the students
into 5~communities, each corresponding to a different grade,
with the exception that all the pupils of one of the classes
of year~4 are assigned to the same community as the pupils
of year~5. This partition, with a modularity of~$0.3066$,
is consistent with younger children tending to separate by
age in their individual interactions, and with such boundaries
becoming less rigid as they grow up. The communities of 3-people
interactions ($k=3$) reveal a similar situation. Year-3 students
keep to themselves, whereas one entire class of year-1 pupils
joins the two of year-2 students to form a community. The
pupils of the last two years are together in the same community.
This division, with a hypermodularity of~$0.71$, is again
consistent with the dynamics just described. Year-3 students
consider the two lower grades too young to form social interactions
with, and they are themselves considered too young by the two higher
grades. The data for $k=4$ show an intriguing division, with
a hypermodularity of~$0.7632$. Years~2 and~3 are joined in
a single community, and years~1, 4 and~5 form another one,
with two individual isolated students. A possible explanation
for the former is that in looser groups of four people it
is easier to include one or two members of different age
without introducing too much disparity of maturity than it
is in tighter groups of~2 or 3~people. The latter community
is perhaps more striking. The age difference between the
individuals in year~1 and those in year~4 and year~5 is
of 3 to 4~years, and this community seems to result from
the presence of siblings, who join the age-based social
group of their brothers and sisters. Indeed, the average
age gap between siblings in the period where the data were
collected was, in fact, $3.5$~years~\cite{Sch20}. Finally,
for~$k=5$ the network is almost completely fragmented,
and the only nodes left with some edges separate into communities
that are either uniform, or with at most one grade of difference
between the members. Note that the network of 5-people
interactions consists of 6~components, 5~of which contain
a single hyperedge, and thus have a vanishing hypermodularity.
The division of the only nontrivial component has a hypermodularity
of~$0.662$.

\begin{figure*}[t]
\centering
\includegraphics[width=0.38\textwidth]{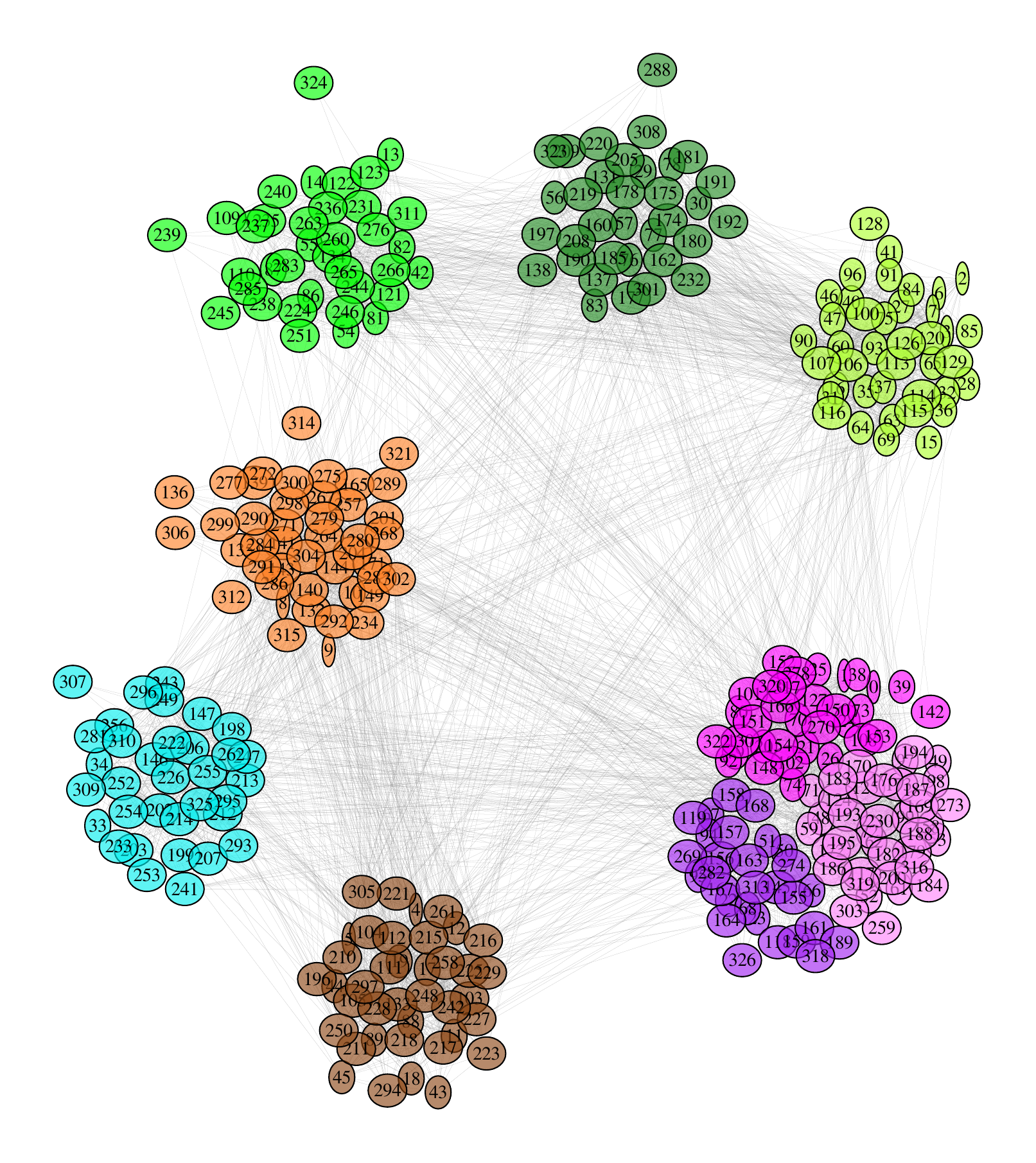}$\qquad\qquad\qquad\qquad\quad$
\includegraphics[width=0.45\textwidth]{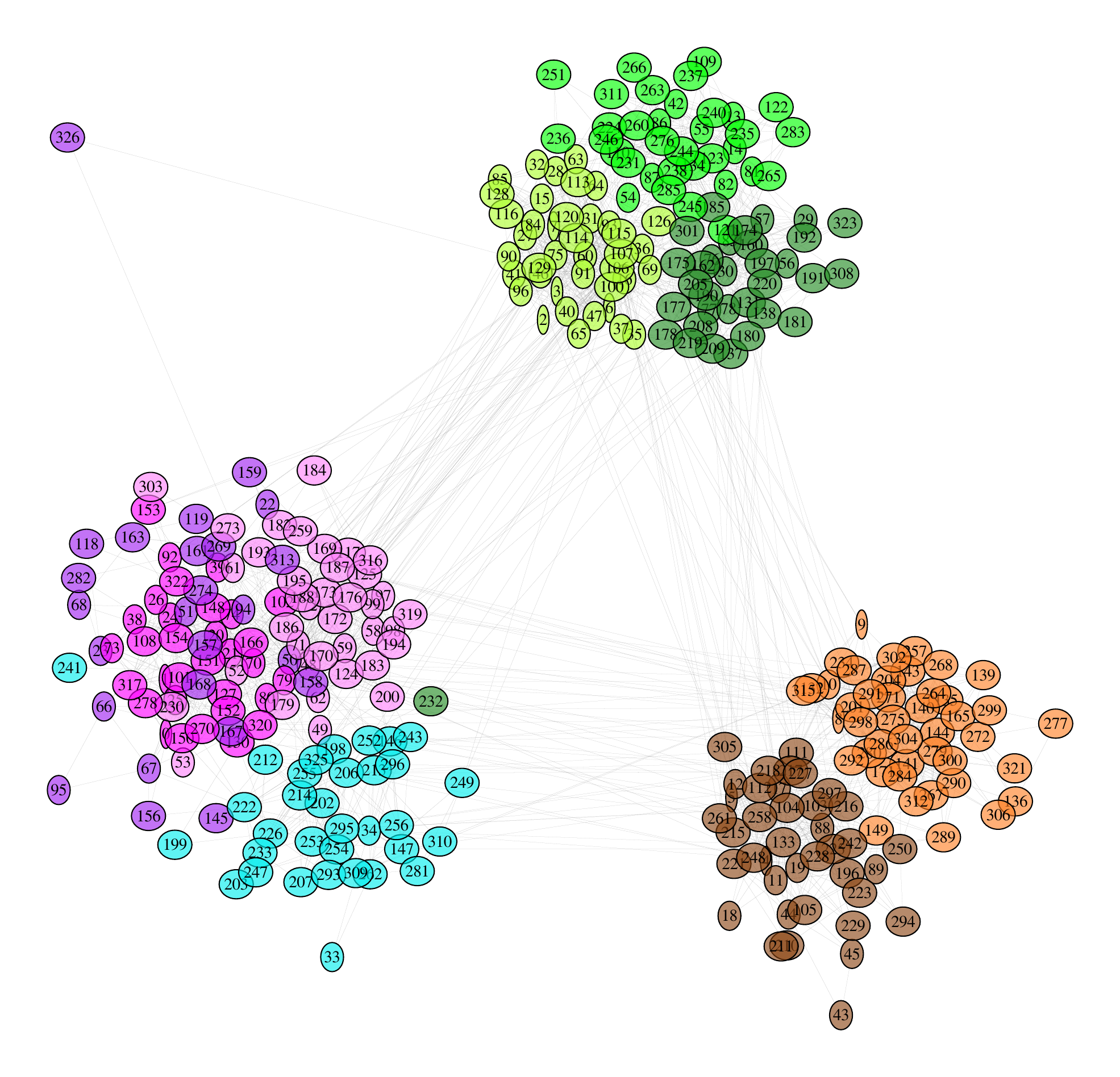}
\includegraphics[width=0.45\textwidth]{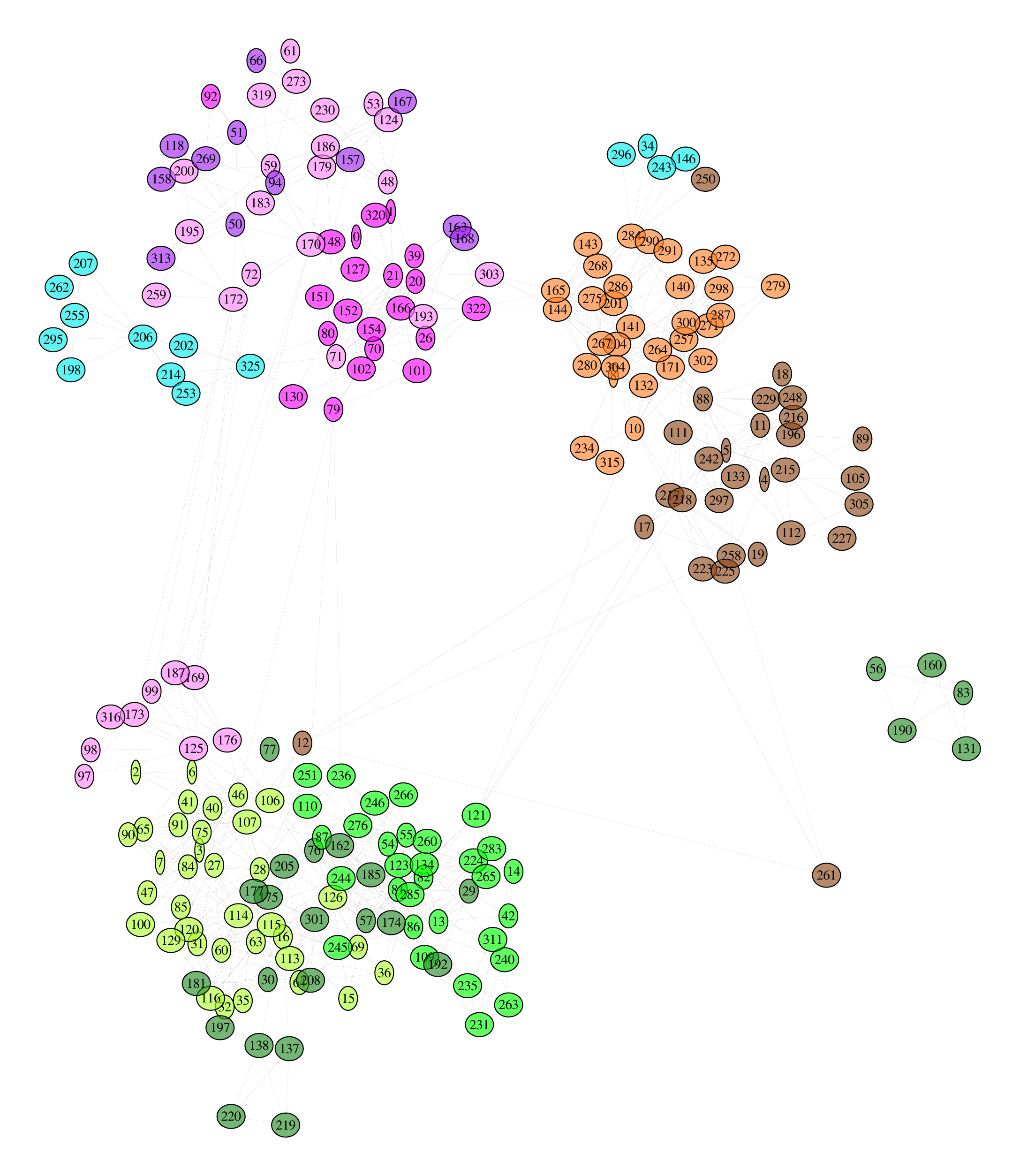}$\qquad\qquad\quad$
\includegraphics[width=0.45\textwidth]{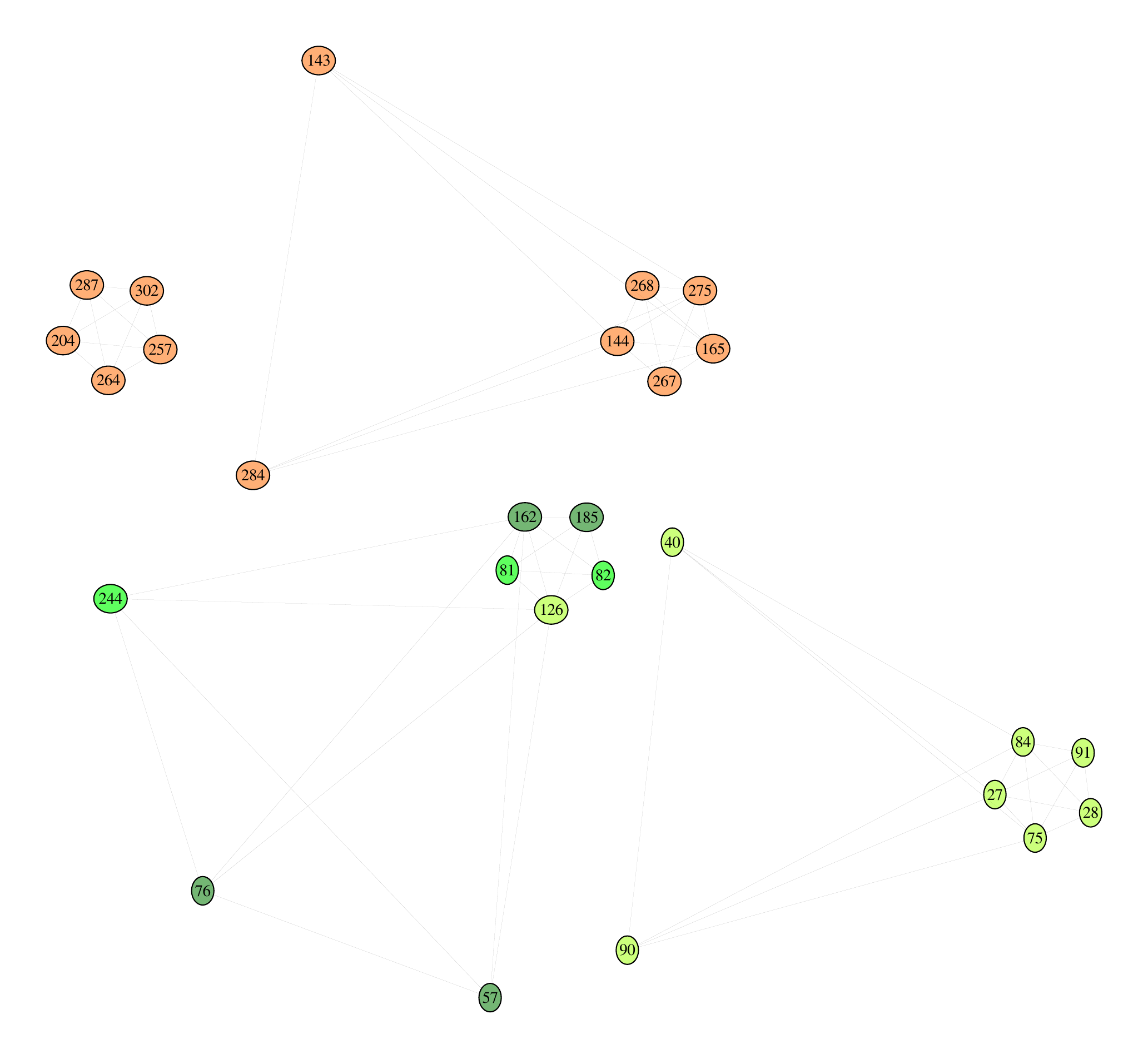}
\caption{\textbf{Higher-order communities of high-school students
provide insights into closeness of interests.} The communities of
dyadic interactions between high-school students (top-left) show
a strong division between topics (colours) and classes (shades),
with the exception of mathematics and physics students (violet),
who interact across classes, challenging the stereotypical opinion
of more mathematically minded people being unable to form meaningful social
contacts. The communities of groups of 3~people (top-right) reveal
the proximity of thought between psychology and mathematics-physics
students, who are now joined in a single community. The 4-people
interactions (bottom-left) extend this insight, revealing how biology
students (green) are substantially isolated, whereas almost all the
other students belong to the same community, with the exception of
a small separate clique. Finally, at $k=5$ the network consists of
only 4~small study groups. In all panels, green nodes correspond
to biology students, violet nodes to mathematics and physics students,
brown nodes to chemistry students and blue nodes to psychology students.
All topics except psychology include more than one class, signified
by the shade of the colour. Note that in preparing the figure, hyperedges
($k>2$) have been projected onto cliques for ease of visualization.}\label{HigSch}
\end{figure*}
The analysis of the high-school data set, shown in Fig.~\ref{HigSch},
also provides insights into the social dynamics of the groups of students.
When one considers only dyadic interactions, the network separates into
communities that are completely identified not only by the topic studied,
but also by the specific class in which the students are enrolled. This
partition has a modularity of~$0.5915$, and the only exception is that
the mathematics and physics students form a single community, rather
than splitting into their three classes. This evidently disproves the
stereotype that sees mathematics and physics students as socially inept.
In fact, in this study, they are the only group of people who socialize
across classes, albeit in the same topic. When we take $k=3$, the network
is very highly modular, with 3~large communities plus an individual
node and a hypermodularity of~$0.8313$. Here, the communities are entirely
topic-based, but the students of psychology are joined with those of
mathematics and physics. A possible explanation for this is that both
physicists and psychologists study complex systems, even though using
different means of representation. Thus, they have a larger affinity
than one may imagine at first thought. At the next larger size of
edges, $k=4$, the meta-group of mathematics-physics-psychology students
joins that of chemistry ones, with the notable exception of 9~mathematics-and-physics
students who are now assigned to the community of biology. Also, the
network starts fragmenting, with an isolated clique of 5~biology students.
This partition of the large component has a hypermodularity of~$0.8527$,
and it is mostly caused by a few students who act as bridges between
disciplines, and whose importance increases as the network has substantially
fewer edges. For example, psychology student~325 is well connected with
other psychology students and with mathematics-physics ones, chemistry student~284
is highly connected with a clique of psychology students and with numerous
other chemistry ones, and mathematics-physics students~125 and~176 have
several links to other students in the same class as well as to biology
ones. Finally, when $k=5$ very few nodes remain with some edges. Five
of them are a clique of chemistry students, and the other three components
consist entirely of students of the same topic (two components for biology
and one for chemistry). This suggests that these components are study
groups, and the larger communities identified within them are their
core members. It should be noted, nonetheless, that only the larger
biology component is reasonably modular, with its partition corresponding
to a hypermodularity of~$0.3319$, whereas both other components have
a hypermodularity of~$0.1723$, which is below what one would expect
from a random hypergraph, as discussed above.

\section{Conclusions}
In this article, we have considered the task of detecting communities
in networks with interactions beyond pairwise ones. Our way to address
this problem is by defining a hypermodularity function for hypergraphs
using a formalism that is conducive to the use of spectral methods to
find the bisection of the network that maximizes it. Specifically, we
have shown that this can be achieved by the use of higher-order singular
value decomposition, if one rewrites the condition that the elements
of a given set of nodes all belong to the same community as a separable
expression of spin-like variables. To do so, we demonstrated a closed
combinatorial expression that allows one to ultimately turn the hypermodularity
equation into a vector form.

After implementing an algorithm based on these ideas,
we applied it to random hypergraphs, finding a physiological
value of maximum hypermodularity that one can expect
to find in networks without an imposed community structure.
This is due to the nonzero probability that, even placing
all edges at random, one can end up with modules of nodes
whose internal edges are denser than their external ones.
Notably, the hypermodularity of random hypergraphs never
exceeded~$0.2$, suggesting that this value can be considered
as a rough threshold to determine the presence of strong
communities.

Note that this result offers a natural explanation
of the often-mentioned issue of ``overfitting'' in
modularity and related measures. The existence of this
problem is usually explained by noting that, occasionally,
networks generated randomly from an Erdős-Rényi ensemble
can admit partitions with nonzero modularity. Indeed,
we have shown that values of hypermodularity as high
as~$0.8$ can be found when the starting network is
not connected. However, we strongly believe that applying
any community detection method to a disconnected network
is a methodological error. Since no connections exist
between components, any communities within the network
will have to be confined each to a single component.
But then, rather than the overall value of the modularity,
one should consider the components individually, and
measure the modularity of each by itself. Additionally,
the rough limit of~$0.2$ for the maximum hypermodularity
of a partition of a random hypergraph can be interpreted
as related to the probability that, placing a number
of edges at random, a set of more densely connected
nodes arises. Also note that this value, and indeed
the entire trend shown in the top panels of Fig.~\ref{hER},
is very much similar to what has been measured in traditional
graphs~\cite{Gui04}, showing that the behaviour of
hypermodularity is consistent with its originating
principles.

In the more than two decades since the seminal article
in Ref.~\cite{Gui04},
these considerations have led to the creation of numerous
methods based on Bayesian paradigms, and most notably on
the use of stochastic block models. These are a class of
models to build random graphs based on the idea of fixing
a number of sets of nodes as communities, and placing
edges internal and external to such modules with some chosen probabilities.
These methods constitute an alternative to other community-detection
approaches. However, they also present notable problems,
such as the tendency to produce poor partitions, sometimes
including disconnected communities, on some networks~\cite{Par25}.

In turn, we do not expect hypermodularity to be exempt
from drawbacks either. In fact, we believe that it is likely
affected by a higher-order version of the resolution
limit~\cite{For06}. However, we also conjecture that the
resolution limit may have a lower importance than it has
on traditional networks, due to the combinatorial factor
weighing the product of degrees in the hypermodularity
equation. In the near future, we plan to investigate
this question in detail and quantitatively evaluate the
limitations of hypermodularity, potentially addressing them
via suitable modifications of its formulation.

By maximizing hypermodularity in its current form, our algorithm
was able to detect communities at different levels of collective
interaction when applied to real-world data.
The communities thus identified can all be explained
by assuming reasonable mechanisms for their formation.
In fact, it can be argued that the communities found
on social contact networks represent similarity of thought
processes and reflect the biases of the individuals.
Thus, we believe that our formulation of hypermodularity
and our algorithm constitute powerful tools for the
analysis of networks with higher-order interactions,
and are highly valuable methods for extracting hidden
information from data sets of diverse nature.

\acknowledgments
The author acknowledges fruitful discussions with Ekaterina
Vasilyeva and Stefano Boccaletti. This work was supported by
the Bulgarian Ministry of Education and Science under project
number BG-RRP-2.004-0006-C02.

\section*{DATA AVAILABILITY}
The data that support the findings of this article are openly available~\cite{figur}.
The code implementing the algorithm described in this article and used
to generate the data is openly available~\cite{code}.

\appendix
\section{Identity of the flattenings}\label{flatident}
To formally show why all the flattenings are identical,
start by explicitly writing the $k$-th standard flattening
of~$\mathbf B$:
\begin{equation}
 \mathbf E^{(k)} = \begin{pmatrix}
                    B_{1,1,\dotsc,1} & B_{2,1,\dotsc,1} & \hdots & B_{N,N,\dotsc,1} \\
                    B_{1,1,\dotsc,2} & B_{2,1,\dotsc,2} & \hdots & B_{N,N,\dotsc,2} \\
                    \vdots           & \vdots           & \ddots & \vdots \\
                    B_{1,1,\dotsc,N} & B_{2,1,\dotsc,N} & \hdots & B_{N,N,\dotsc,N}
                   \end{pmatrix}\:.
\end{equation}
This equation shows that the row index of~$\mathbf E^{(k)}$
corresponds to the $k$-th index of~$\mathbf B$. Also, the
column index of~$\mathbf E^{(k)}$, which numerically ranges
from~1 to~$N^{k-1}$, is equal to the positional number of the
corresponding choice of values for the remaining indices of~$\mathbf B$
in an inverse lexicographic order. In other words, each element~$E^{(k)}_{i,j}$
corresponds to the element of~$\mathbf B$ whose $k$-th index
is equal to~$i$ and whose other indices $v_1,\dotsc,v_{k-1}$
satisfy the equation
\begin{multline}
 j = v_1 N^0 + \lp v_2-1\rp N^1 + \lp v_3-1\rp N^2\\
 + \dotsb + \lp v_{k-1} -1\rp N^{k-2}\:.
\end{multline}
Effectively, the equation above provides the unique
\mbox{$k-1$}~digits of the column numbers of the elements
of~$\mathbf E^{(k)}$ expressed in a base-$N$ positional
system.

Now consider the generic $x$-th standard flattening
of~$\mathbf B$:
\begin{widetext}
 \begin{equation}
  \mathbf E^{(x)} = \begin{pmatrix}
                     B_{1,1,\dotsc,1,1,1,\dotsc,1} & B_{2,1,\dotsc,1,1,1,\dotsc,1} & \hdots & B_{N,N,\dotsc,N,1,N,\dotsc,N} \\
                     B_{1,1,\dotsc,1,2,1,\dotsc,1} & B_{2,1,\dotsc,1,2,1,\dotsc,1} & \hdots & B_{N,N,\dotsc,N,2,N,\dotsc,N} \\
                     \vdots                        & \vdots                        & \ddots & \vdots \\
                     B_{1,1,\dotsc,1,N,1,\dotsc,1} & B_{2,1,\dotsc,1,N,1,\dotsc,1} & \hdots & B_{N,N,\dotsc,N,N,N,\dotsc,N}
                    \end{pmatrix}\:.
 \end{equation}
\end{widetext}
Here, the row index of~$\mathbf E^{(x)}$, which
is the index of~$\mathbf B$ that changes between
the elements in each column, is the \mbox{$x$-th}
one. So, given an element~$E^{(x)}_{i,j}$, it corresponds
to the element of~$\mathbf B$ whose $x$-th index
is equal to~$i$ and whose other indices are such
that
\begin{multline}
 j = v_1 N^0 + \lp v_2-1\rp N^1 + \dotsb + \lp v_{x-1}-1\rp N^{x-2} +\\
 \lp v_{x+1}-1\rp N^{x-1} + \dotsb + \lp v_k -1\rp N^{k-2}\:.
\end{multline}

Then, fix a row index~$l$ and a column
index~$m$ for the two flattenings, and
take their elements~$E^{(k)}_{l,m}$ and~$E^{(x)}_{l,m}$.
The former maps to the element of the
data tensor~$B_{v_1,v_2,\dotsc,v_{x-1},v_x,v_{x+1},\dotsc,v_{k-1},l}$
whose last index is equal to~$l$ and whose
other indices correspond to the unique
decomposition of~$m$ in powers of~$N$.
The other, in the same way, corresponds
instead to~$B_{v_1,v_2,\dotsc,v_{x-1},l,v_{x+1},\dotsc,v_k}$.
Thus, the indices of the latter element
of~$\mathbf B$ are a permutation of those
of the former. But, since~$\mathbf B$
is hypersymmetric, $B_{v_1,v_2,\dotsc,v_{x-1},v_x,v_{x+1},\dotsc,v_{k-1},l}=B_{v_1,v_2,\dotsc,v_{x-1},l,v_{x+1},\dotsc,v_k}$,
which, in turn, means that, for any choice
of~$l$ and~$m$, $E^{(x)}_{l,m}=E^{(k)}_{l,m}$
for all~$x$. Thus, all the standard flattenings
of the data tensor of a simple hypergraph
are identical, and in the following we
use the $k$-th ones without loss of generality,
omitting explicit mention of the dimension
used for their construction.

\section{Building a vector equation}\label{vectoreq}
To turn Eq.~\eqref{hypermod2} into a vector equation,
first use Eq.~\eqref{deltas} to explicitly rewrite
the product chain for $k=3$:
\begin{equation}
 \begin{split}
  \delta_{C_{v_1}, C_{v_2}} \delta_{C_{v_2}, C_{v_3}} &= \frac{1}{2}\lp s_1 s_2 + 1\rp\frac{1}{2}\lp s_2 s_3 + 1\rp\\
  &= \frac{1}{4}\lp s_1 s_2^2 s_3 + s_1 s_2 + s_2 s_3 + 1\rp\:.
 \end{split}
\end{equation}
Note that in the equation above, $s_1$, $s_2$
and~$s_3$ indicate the community assignments
of the nodes chosen as~$v_1$, $v_2$ and~$v_3$,
and \emph{not} those of nodes~1, 2 and~3. Thus,
for~$k=3$, Eq.~\eqref{hypermod2} becomes
\begin{equation}
 \begin{split}
  Q &= \frac{1}{3!m}{\sum}_{\lbr v_1, v_2, v_3\rbr}\lp B_{v_1,v_2,v_3}\prod_{\substack{i,j=1\\i\neq j}}^3 \delta_{C_{v_i}, C_{v_j}}\rp\\
  &= \frac{1}{3!m}{\sum}_{\lbr v_1, v_2, v_3\rbr}\ls B_{v_1,v_2,v_3} \frac{1}{4}\lp s_1 s_2^2 s_3 + s_1 s_2\right.\right.\\
  &\quad + s_2 s_3 + 1\big)\bigg]\\
  &= \frac{1}{3!m}{\sum}_{\lbr v_1, v_2, v_3\rbr}\ls B_{v_1,v_2,v_3} \frac{1}{4}\lp s_1 s_3 + s_1 s_2\right.\right.\\
  &\quad + s_2 s_3 + 1)\bigg]\:,
 \end{split}
\end{equation}
where, in the last line, we have used~$s_i^2=1$.
But the sum above runs over all combinations of
3~nodes, which means that all the elements of~$\mathbf B$
are considered. In turn, this means that, since
the sum of all the elements of~$\mathbf B$ vanishes
by construction, we can drop any constant term from
the equation, obtaining
\begin{widetext}
\begin{equation}
 \begin{split}
  Q &= \frac{1}{3!m}{\sum}_{\lbr v_1, v_2, v_3\rbr}\ls B_{v_1,v_2,v_3} \frac{1}{4}\lp s_1 s_3 + s_1 s_2 + s_2 s_3 \rp\rs\\
  &= \frac{1}{3!m}{\sum}_{\lbr v_1, v_2, v_3\rbr}\lbr B_{v_1,v_2,v_3} \frac{1}{4}\ls s_3\lp s_1 + s_2\rp + s_1 s_2 \rs\rbr\:.
 \end{split}
\end{equation}
\end{widetext}
But, again because of the sum and of the nature
of~$\mathbf{B}$, the index of~$s_1$ in the last
term above is a dummy index. Thus, we can change
that~$s_1$ to~$s_3$, and get
\begin{equation}
 Q = \frac{1}{3!m}{\sum}_{\lbr v_1, v_2, v_3\rbr}\ls B_{v_1,v_2,v_3} \frac{1}{4} s_3\lp s_1 + 2 s_2\rp\rs\:,
\end{equation}
which we rewrite as
\begin{equation}
 Q = \frac{1}{3!m}{\sum}_{\lbr v_1, v_2, v_3\rbr}B_{v_1,v_2,v_3} S^{(3)}\:,
\end{equation}
where we have put
\begin{equation}\label{S3}
 S^{(3)}=\frac{1}{4} s_3\lp s_1 + 2 s_2\rp\:.
\end{equation}
This equation conveniently separates the three spin-like
variables into a product of one of them and an expression
featuring the other two. Thus, the equation for hypermodularity
can be rewritten as
\begin{equation}\label{vechypmodk3}
 Q = \frac{1}{4\cdot 3!m}\mathbf s^{\mathrm T}\mathbf E\boldsymbol\sigma^{(3)}\:,
\end{equation}
where~$\mathbf s$ is a vector containing
the community assignments of all nodes,
and~$\boldsymbol\sigma^{(3)}$ is an $N^2$-dimensional
vector whose elements are the values of
the expression $s_1+2s_2$ computed for all
possible pairs of nodes in the same order
as those corresponding to the columns of~$\mathbf E$.

To find a general-use formula, repeat
the calculation for $k=4$, by writing
the chain product of deltas in terms
of spin-like variables, dropping any
constants and replacing dummy indices
with the 4th one:
\begin{equation}\label{S4}
 \begin{split}
  S^{(4)} &= \delta_{C_{v_1}, C_{v_2}} \delta_{C_{v_2}, C_{v_3}} \delta_{C_{v_3}, C_{v_4}}\\
  &= \frac{1}{2}\lp s_1 s_2 + 1\rp\frac{1}{2}\lp s_2 s_3 + 1\rp\frac{1}{2}\lp s_3 s_4 + 1\rp\\
  &= \frac{1}{8} \lp s_1s_2^2s_3^2s_4 + s_1s_2^2s_3 + s_1s_2s_3s_4 + s_1s_2\right.\\
  &\quad + \left. s_2s_3^2s_4 + s_2s_3 + s_3s_4 + 1\rp\\
  &= \frac{1}{8} \ls s_4\lp s_1+s_2+s_3+s_1s_2s_3\rp\right. \\
  &\quad + \left. s_1s_2 + s_1s_3 + s_2s_3 + 1\rs\\
  &\rightarrow \frac{1}{8} \ls s_4\lp s_1+s_2+s_3+s_1s_2s_3\rp + s_1s_2 + s_1s_3 + s_2s_3 \rs\\
  &\rightarrow \frac{1}{8} \ls s_4\lp s_1+s_2+s_3+s_1s_2s_3\rp + s_4s_2 + s_4s_3 + s_4s_3 \rs\\
  &\rightarrow \frac{1}{8} s_4 \lp s_1 + 2s_2 + 3s_3 + s_1s_2s_3\rp\:,
 \end{split}
\end{equation}
where we have used a right arrow, rather
than the equal sign, to indicate functional,
but not formal, equality.
Similarly, for $k=5$, one gets
\begin{multline}
 S^{(5)} = \frac{1}{16} s_5 \lp s_1 + 2s_2 + 3s_3 + 4s_4 + s_1s_2s_3 \right.\\
 \left. + s_1s_2s_4 + s_1s_3s_4 + 2s_2s_3s_4\rp\:,
\end{multline}
and for $k=6$ it is
\begin{multline}\label{S6}
 S^{(6)} = \frac{1}{32} s_6 \lp s_1 + 2s_2 + 3s_3 + 4s_4 + 5s_5 + s_1s_2s_3 \right.\\
 + s_1s_2s_4 + s_1s_2s_5 + s_1s_3s_4 + s_1s_3s_5 + s_1s_4s_5 + 2s_2s_3s_4 \\
 \left. + 2s_2s_3s_5 + 2s_2s_4s_5 + 3s_3s_4s_5 + s_1s_2s_3s_4s_5 \rp\:.
\end{multline}

It becomes then clear that the vector expression
for the hypermodularity of a bipartition, which
we wrote in Eq.~\eqref{vechypmodk3} for $k=3$, can
be generalized to any~$k$ in the form given by Eqs.~\eqref{vechypmod}
and~\eqref{combvec}, which we rewrite here for the
sake of completeness:
\begin{widetext}
\begin{gather}
 Q = \frac{1}{2^{k-1}k!m}\mathbf s^{\mathrm T}\mathbf E\boldsymbol\sigma^{(k)}\\
 \sigma^{(k)}_{\xi_1,\xi_2,\dotsc,\xi_{k-1}} = \sum_{\substack{r=1\\r\text{ odd}}}^{k-1}\left.\sum_{\substack{\text{ordered $r$-choices }\lbr\varsigma_1,\varsigma_2,\dotsc,\varsigma_r\rbr\\\text{from }\lbr\zeta_1,\zeta_2,\dotsc,\zeta_{k-1}\rbr}}\alpha_1\varsigma_1\varsigma_2\dotsb\varsigma_r\right|_{\zeta_1=s_{\xi_1}, \zeta_2=s_{\xi_2}, \dotsc, \zeta_{k-1}=s_{\xi_{k-1}}}\:,
\end{gather}
\end{widetext}
where~$\alpha_1$ is the ordinal index
of~$\varsigma_1$, so that $\alpha_1=1$
if $\varsigma_1=\zeta_1$, $\alpha_1=2$
if $\varsigma_1=\zeta_2$, and so on.

Note that this demonstration not only
offers a straightforward way to employ
spectral methods in search of the best
partition of a higher-order network into
two communities, but it also shows that
in classification tasks, the assignment
corresponding to the first higher-order
singular vector of a data tensor is indeed
the best one, and not just an approximation
of it, in turn providing an explanation
of the success of methods based on higher-order
SVD in machine learning.

\section{Repeated bisections}\label{repbisec}
To find the expression for the correction
to use when repeating the bisection process,
start from Eq.~\eqref{deltaQ}, which we rewrite
here for convenience:
\begin{multline}
 \Delta Q=\frac{1}{k!m}\lp\sum_{\text{$k$-sets in $C$}}B_{v_1,\dotsc,v_k}\delta_{C_{v_1},C_{v_2}}\dotsb\delta_{C_{v_{k-1}},C_{v_k}}\right.\\
 \left.- \sum_{\text{$k$-sets in $C$}}B_{v_1,\dotsc,v_k}\rp\:.
\end{multline}

Now, write the product of deltas in terms of the spin-like
variables. However, note that, for the reasons discussed in
Subsection~\ref{specpar}, the constant term can no longer
be neglected.
This results in
\begin{widetext}
 \begin{equation}
  \Delta Q=\frac{1}{k!m}\ls\frac{1}{2^{k-1}}{\sum}_{s_{v_1}}\dotsb{\sum}_{s_{v_k}}B_{v_1,\dotsc,v_k}\lp s_{v_k} f(s_{v_1},\dotsc,s_{v_{k-1}})+1\rp - {\sum}_{s_1}\dotsb{\sum}_{s_{v_k}}B_{v_1,\dotsc,v_k}\rs\:,
 \end{equation}
\end{widetext}
where we have explicitly expanded the sums and~$f$
is a polynomial in the first $k-1$~spin-like variables,
as in Eq.~\eqref{S3} and in Eqs.~(\ref{S4}--\ref{S6}).

\begin{figure*}[t]
\centering
\includegraphics[width=0.75\textwidth]{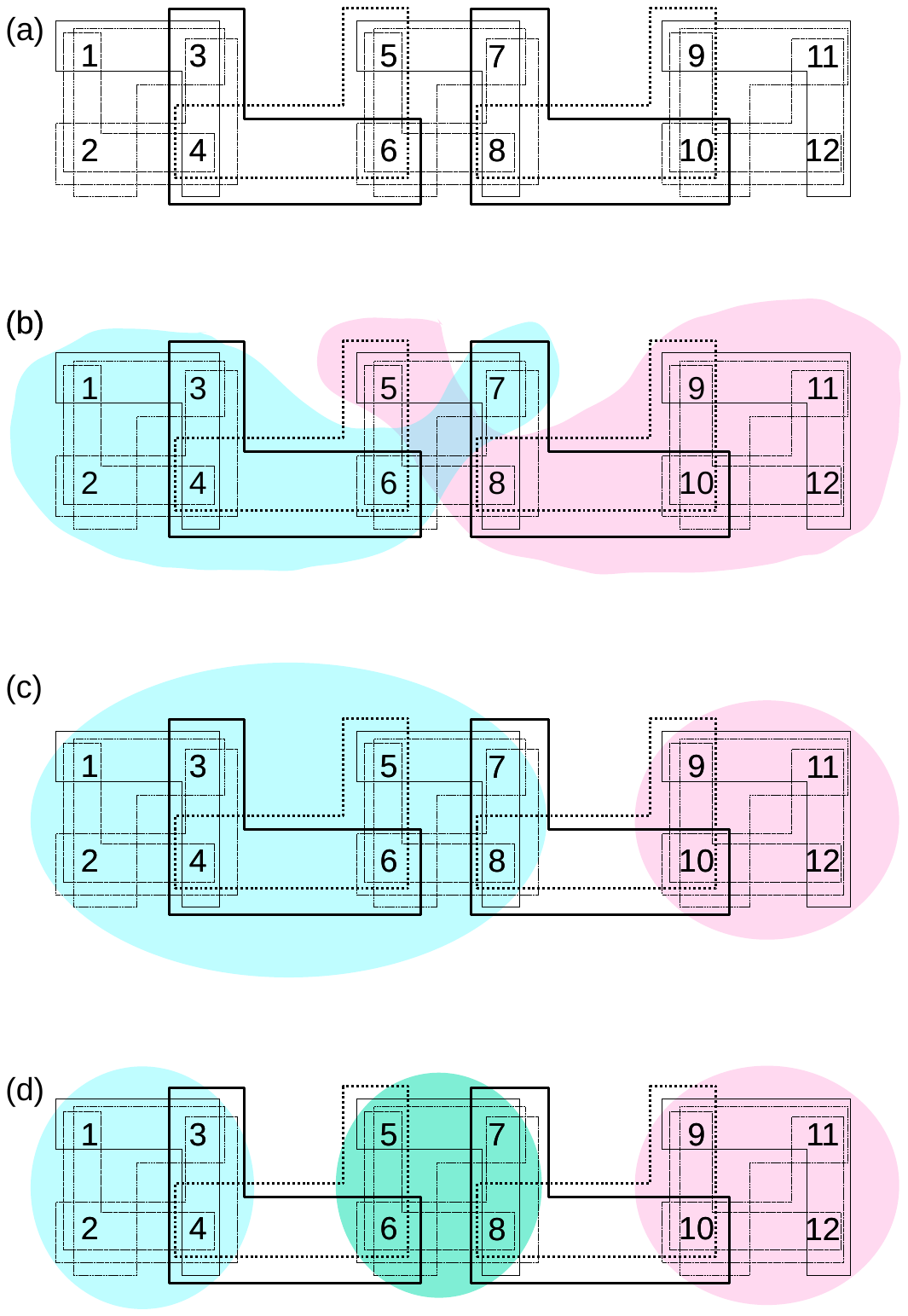}
\caption{\textbf{Worked example of the hypermodularity maximization
algorithm.} (a) The network is a 12-node 3-uniform hypergraph consisting
of three cliques, specifically on nodes~1--5, 5--8 and 9--12, joined
by edges~$(3,4,7)$, $(4,5,7)$, $(6,8,11)$ and~$(8,9,11)$. All the edges
within each clique are shown here as lines of different style. The
edges that join the cliques are slightly thicker than the others. (b)
After the first bisection, the best partition found consists of two
communities, shown here in blue and pink, with a hypermodularity of~$0.375$.
(c) After the first local node-level refinement, the communities change,
and the hypermodularity rises to~$0.519531$. (d) Further passes of
the method split the blue community into two, shown here in blue and
emerald~green. At this point, no further improvement is possible, and
the algorithm stops. The hypermodularity of the final partition is~$0.63623$.}\label{figex}
\end{figure*}
Then, expanding the product, we obtain
\begin{widetext}
 \begin{equation}
  \begin{split}
   \Delta Q &= \frac{1}{k!m}\lp\frac{1}{2^{k-1}}{\sum}_{s_{v_1}}\dotsb{\sum}_{s_{v_k}}B_{v_1,\dotsc,v_k} s_{v_k} f(s_{v_1},\dotsc,s_{v_{k-1}}) +\frac{1}{2^{k-1}} {\sum}_{s_{v_1}}\dotsb{\sum}_{s_{v_k}}B_{v_1,\dotsc,v_k} - {\sum}_{s_{v_1}}\dotsb{\sum}_{s_{v_k}}B_{v_1,\dotsc,v_k}\rp\\
   &= \frac{1}{k!m}\lp\frac{1}{2^{k-1}}{\sum}_{s_{v_1}}\dotsb{\sum}_{s_{v_k}}B_{v_1,\dotsc,v_k} s_{v_k} f(s_{v_1},\dotsc,s_{v_{k-1}})-\frac{2^{k-1}-1}{2^{k-1}} {\sum}_{s_{v_1}}\dotsb{\sum}_{s_{v_k}}B_{v_1,\dotsc,v_k}\rp\\
   &= \frac{1}{2^{k-1}k!m}\ls{\sum}_{s_{v_1}}\dotsb{\sum}_{s_{v_k}}B_{v_1,\dotsc,v_k} s_{v_k} f(s_{v_1},\dotsc,s_{v_{k-1}})-\lp 2^{k-1}-1\rp {\sum}_{s_{v_1}}\dotsb{\sum}_{s_{v_k}}B_{v_1,\dotsc,v_k}\rs\\
   &= \frac{1}{2^{k-1}k!m}\ls{\sum}_{s_{v_1}}\dotsb{\sum}_{s_{v_k}}B_{v_1,\dotsc,v_k} s_{v_k} f(s_{v_1},\dotsc,s_{v_{k-1}})\right.\\
   &\quad \left. -\lp 2^{k-1}-1\rp {\sum}_{s_{v_1}}\dotsb{\sum}_{s_{v_k}}\delta_{v_1,v_2}\dotsb\delta_{v_{k-1},v_k}{\sum}_{s_{w_1}}\dotsb{\sum}_{s_{w_{k-1}}}B_{w_1,\dotsc,w_{k-1},v_k}\rs\\
   &= \frac{1}{2^{k-1}k!m}\ls{\sum}_{s_{v_1}}\dotsb{\sum}_{s_{v_k}}B_{v_1,\dotsc,v_k} s_{v_k} f(s_{v_1},\dotsc,s_{v_{k-1}})\right.\\
   &\quad \left. -\lp 2^{k-1}-1\rp {\sum}_{s_{v_1}}\dotsb{\sum}_{s_{v_k}}\delta_{v_1,v_2}\dotsb\delta_{v_{k-1},v_k}{\sum}_{s_{w_1}}\dotsb{\sum}_{s_{w_{k-1}}}B_{w_1,\dotsc,w_{k-1},v_k}\frac{s_{v_k}f(s_{v_1},\dotsc,s_{v_{k-1}})}{2^{k-1}-1}\rs\:,
  \end{split}
 \end{equation}
\end{widetext}
where, in the last line, we have used
the fact that $s_{v_k}f(s_{v_1},\dotsc,s_{v_{k-1}})=2^{k-1}-1$
when $v_1=v_2=\dotsb=v_k$. But then,
it is
\begin{widetext}
 \begin{equation}
  \Delta Q = \frac{1}{2^{k-1}k!m}{\sum}_{s_{v_1}}\dotsb{\sum}_{s_{v_k}}s_{v_k}\lp B_{v_1,\dotsc,v_k} - \delta_{v_1,v_2}\dotsb\delta_{v_{k-1},v_k}{\sum}_{s_{w_1}}\dotsb{\sum}_{s_{w_{k-1}}}B_{w_1,\dotsc,w_{k-1},v_k}\rp f(s_{v_1},\dotsc,s_{v_{k-1}})\:,
 \end{equation}
\end{widetext}
which means that we can express the change
in hypermodularity as in Eqs.~\eqref{repbis}
and~\eqref{correq}, which we rewrite below:
\begin{widetext}
\begin{gather}
 \Delta Q = \frac{1}{2^{k-1}k!m}\mathbf s^{\mathrm T}\mathbf{E'}\boldsymbol\sigma^{(k)}\\
  B'_{v_1,\dotsc,v_k} = \begin{cases}
                        B_{v_1,\dotsc,v_k} - {\sum}_{s_{w_1}}\dotsb{\sum}_{s_{w_{k-1}}}B_{w_1,\dotsc,w_{k-1},v_k} & \quad\text{if $v_1=v_2=\dotsb=v_k$}\\
                        B_{v_1,\dotsc,v_k} & \quad\text{otherwise.}
                       \end{cases}
\end{gather}
\end{widetext}

\section{A worked example}\label{example}
Here, we provide an example of how our method works,
by going step-by-step through the partitioning of a
12-node 3-uniform hypergraph. The hypergraph consists
of three groups of nodes, namely nodes~1--4, 5--8 and
9--12, with all possible hyperedges of size~3 within
each group. Additionally, as illustrated in Fig.~\ref{figex}~(a),
edges $(3,4,7)$, $(4,5,7)$, $(6,8,11)$ and~$(8,9,11)$
join the cliques.

In the beginning, all the nodes belong to the same
community, and the hypermodularity is~0. Computing
the left-singular vector corresponding to the largest
singular value results in a partition where nodes~1,
2, 3, 4, 6 and~7 are assigned to community~1, and
the remaining ones are assigned to community~2. Since this
partition, depicted in Fig.~\ref{figex}~(b), has a
hypermodularity of~$0.375$, it is accepted.

Next, a local node-level refinement is attempted.
The best set of moves consists in shifting nodes~5
and~8 from community~2 to community~1. These moves
yield an increase in hypermodularity of~$0.144531$.
Thus, they are accepted, resulting in the partition
shown in Fig.~\ref{figex}~(c), with hypermodularity
of~$0.519531$.

No global node-level refinement is possible,
since it would be just a repetition of the last
refinement step. Similarly, joining the two
communities would make the hypermodularity vanish,
and therefore no further modifications happen
at this stage.

Trying to bisect community~1 using the higher-order
SVD does not yield an increase in hypermodularity.
Therefore, no bisection is performed on it. However,
the best result from the local node-level refinement
step consists in moving nodes~5--8 to a new community,
which we label community~3. This move yields an increase
of hypermodularity of~$0.116699$. Thus, it is accepted,
and the resulting partition, shown in Fig.~\ref{figex}~(d),
has hypermodularity of~$0.63623$.

Next, a bisection is attempted on community~2,
but this would not result in an increase of hypermodularity,
and therefore none is performed. Similarly, local
and global node-level refinements and community-joining
attempts all would decrease the value of hypermodularity,
and no such moves are accepted. At this point,
community~2 is marked as blocked, since no further
improvement can be obtained by working solely
on it.

Similar attempts at continuing the process,
first on community~1 and then on community~3,
do not produce any improvement. Thus, the whole
procedure stops. The final result is that the
method correctly split the original network
into the three densely-connected modules, producing
a partition with a significant value of hypermodularity.

\end{document}